\documentclass{aa}  

\usepackage{graphicx}
\usepackage{txfonts}
\usepackage{lipsum}
\usepackage{subcaption}         
\usepackage{lscape}             
\usepackage{placeins}           

\usepackage[dvipsnames]{xcolor}
\usepackage{hyperref}
\hypersetup{pdfborder = {0 0 0},
    colorlinks = true,
    linkbordercolor = {white},
    citecolor=NavyBlue,
    urlcolor=Green
} 
\usepackage{xspace}
\newcommand{\numax}{\ensuremath{\nu_{\rm max}}\xspace}                                

\begin{document}

   \title{Ensemble asteroseismology}
   \subtitle{An ensemble approach to detecting signatures of solar-like oscillations in K dwarfs}
   \titlerunning{Ensemble asteroseismology of K dwarfs}

%

   \author{William J. Chaplin\inst{1}\fnmsep\thanks{Corresponding author: w.j.chaplin@bham.ac.uk}
        \and Tiago L. Campante\inst{2,3}
        \and Mikkel N. Lund\inst{4}
        \and Martin B. Nielsen\inst{1}
        \and Guy R. Davies\inst{1}
        \and Emily Hatt\inst{1}
        \and Rachel Howe\inst{1}
        \and Amalie Stokholm\inst{1}
        }

   \institute{School of Physics and Astronomy, University of Birmingham, Birmingham, B15 2TT, United Kingdom
    \and
    Instituto de Astrof\'{\i}sica e Ci\^{e}ncias do Espa\c{c}o, Universidade do Porto,  Rua das Estrelas, 4150-762 Porto, Portugal
    \and
    Departamento de F\'{\i}sica e Astronomia, Faculdade de Ci\^{e}ncias da Universidade do Porto, Rua do Campo Alegre, s/n, 4169-007 Porto, Portugal
    \and
    Stellar Astrophysics Centre, Department of Physics and Astronomy, Aarhus University, Ny Munkegade 120, DK-8000 Aarhus C, Denmark
    }
  
   \authorrunning{Chaplin et al.}
   \date{Received XX, YY}

 
  \abstract
   {Solar-like oscillations have to date been observed in hundreds of main-sequence and sub-giant stars. However, only a select number of detections have been made in K-type dwarfs, using extreme-precision ground-based radial velocity observations and space-based photometric observations made by the NASA \emph{Kepler} and TESS missions. Whilst the upcoming ESA PLATO Mission promises to add to these individual detections, it will do so only in a similar, modest number of stars. }
   {Here, we propose a new ensemble strategy to exploit the PLATO data, in which frequency power spectra on hundreds of K dwarfs lying in constrained ranges of effective temperature are combined in a weighted manner to significantly improve the detectability of the oscillations. Whilst this approach means it is not possible to extract usable constraints on individual oscillation frequencies, it provides a way to detect and measure the characteristics of the composite envelope of oscillation power given by the ensemble, which in turn provides diagnostics of granulation and magneto-convection and the impact of magnetic activity on the modes.} 
   {We used data in the PLATO Input Catalogue (PIC) to make discrete numerical predictions of the detectability of the ensemble spectra. We also derived a simple analytical approximation of our method that obviates the need to perform numerical calculations over a discrete sample of targets, and which serves as a useful tool to make quick predictions for other future or planned missions.}
   {Our predictions indicate that PLATO has the potential to provide solid ensemble detections well into the K-dwarf regime. In summary, PLATO offers an ideal opportunity to exploit this new approach.}
   {}


   \maketitle
   \nolinenumbers

\section{Introduction}
\label{sec:intro}

Thanks to a now extensive database of space- and ground-based observations made, respectively, in photometry and Doppler velocity, solar-like oscillations have been detected in hundreds of main-sequence stars. Detecting oscillations in K dwarfs has, however, proven to be extremely challenging, since the maximum amplitudes of solar-like oscillations decrease as one moves to cooler temperatures on the lower main sequence. At the time of writing, only four detections have been made for K dwarfs using space-based data (e.g. see \citealt{Barclay2013, Campante2016, Hon2024, LundLum2025}). All four detections lie at the boundary between late G and early K-type stars. The detection boundary has been extended into the mid-K range thanks to a concerted effort to utilise extreme-precision ground-based radial velocity capabilities at a number of observatories (e.g. see \citealt{Campante2024, Lundkvist2024, Hon2024, Li2025, Campante2026}). However, even with these new detections, the combined sample of K-dwarf detections still numbers only around ten stars.

Here, we propose a new ensemble strategy that will significantly improve the prospects of detecting signatures of solar-like oscillations in these stars. The basic approach we consider is to co-add the frequency power spectra of many individual stars to reduce the variance of the noise in the combined spectrum compared to that of any individual spectrum, and hence to improve the effective signal-to-noise ratio (S/N) in the oscillations. We leverage the fact that by selecting stars on the lower main sequence from sufficiently narrow ranges in $T_{\rm eff}$, we may assume that to the first order they take similar fundamental properties (i.e. mass, $M$, radius, $R$, and luminosity, $L$). This in turn means that we should expect their oscillation spectra to have similar frequencies of the maximum oscillation power, $\nu_{\rm max}$, and to therefore contribute to a cumulative excess of mode power in a constrained range of frequencies in the combined frequency-power spectrum. Whilst combining power spectra in this way rules out extracting or utilizing individual oscillation frequencies -- the ensemble yields a composite, interleaved spectrum -- it nevertheless opens the way to detecting and measuring the properties of the composite excess of mode power. 

The maximum amplitudes of solar-like oscillations are often described in terms of a dependence on stellar mass and luminosity that goes like $(L/M)^s$ (e.g. see \citealt{Kjeldsen1995, Houdek1999, Samadi2005, Samadi2007}), where $s$ is an exponent that controls the dependence. The lower the value of $s$, the less steep the fall-off in amplitude as one progresses down the lower main sequence. While the above-mentioned ground-based sample is small, it has nevertheless demonstrated that the maximum mode amplitudes appear to fall off more steeply than previously expected as one moves from the G- into the K-dwarf regime, with a breakpoint in behaviour at the transition between classes. While results for G-type stars are consistent with $s \simeq 1$, the exponent appears to be closer to 1.5 for K dwarfs (\citealt{Campante2026}). The ensemble approach offers the prospect of being able to test this dependence over a much larger statistical sample of stars, and further into the K-dwarf regime. By constructing different combinations -- for example, from ranges in $T_{\rm eff}$, measures of activity, and so on -- one can in principle explore the dependence of the amplitudes on stellar properties, and provide new information on the underlying mode physics, granulation, and magnetoconvection (since near-surface turbulence excites and intrinsically damps the modes), and the impact of magnetic activity on these processes. 

An ensemble approach of this type, using a large statistical sample of stars, is now possible thanks to the upcoming ESA PLATO Mission (\citealt{Rauer2025}). PLATO promises to add more individual detections of solar-like oscillations in K dwarfs, but only in modest numbers at the G to K dwarf boundary (\citealt{Goupil2024}). However, its large field of view (much larger than \emph{Kepler}) and high-precision photometry (much lower noise than TESS) make it extremely well suited for exploiting the ensemble method. PLATO will be able to deliver the data required to apply the method, i.e. high-cadence, long-duration light curves for a large enough sample of bright K dwarfs, which neither \emph{Kepler} nor TESS could or can provide. Our aim in this paper is therefore not only to introduce the approach but also to show that it has the potential to deliver positive detections using PLATO data. Our PLATO predictions use the actual targets in the PLATO Input Catalogue (PIC; see \citealt{Montalto2021}), coupled to the best estimates available of the PLATO noise performance. In addition, we also provide simple analytical formulae that require only basic inputs and that may be used to make general predictions for other planned or future missions.

The layout of the rest of the paper is as follows. In Sect.~\ref{sec:comb} we introduce the ensemble approach, and the need to weight the combination of power spectra in an appropriate manner. We also present the weighted summation that describes the standard deviation of the background power spectral density of the combination for the case of a discrete sample of targets. In Sect.~\ref{sec:predan} we derive a simple analytical approximation that obviates the need to perform numerical calculations over a discrete target sample. Section~\ref{sec:detect} explains in detail how to calculate the detection significance and an equivalent detection probability for the weighted combination using the standard deviation, for both the discrete and analytical cases. We then go on to present predictions for PLATO in Sect.~\ref{sec:pred}. Section~\ref{sec:data} introduces the input data from the PIC on which the discrete numerical predictions are based, and Sect.~\ref{sec:params} then explains how parameters needed for the predictions were calculated from the PIC data. The results of our PLATO predictions are presented in Sect.~\ref{sec:res}. We finish in Sect.~\ref{sec:conc} with key conclusions and a discussion.

\section{The ensemble spectrum}
\label{sec:comb}

\subsection{Basic approach}
\label{sec:comb1}

Consider time-domain observations made on a regular cadence, $\Delta t$, that cover a total duration, $T$. In what follows, we assume that they are photometric data, since we focus on predictions for PLATO. We consider power spectra computed from these data are sampled at critical frequencies separated by $\Delta_T = 1/T$, so that the individual bins of the spectra are statistically independent. The observed power spectral density in each bin is distributed about the underlying limit spectrum -- i.e. the spectrum that would be given by an infinite number of independent noise realizations -- with $\chi^2$ two-degrees-of-freedom statistics. The standard deviation of the power within any given bin then corresponds to the underlying limit power for that bin.

Now, assume that we co-add independent spectra, each having the same underlying limit spectrum and noise characteristics. The standard deviation of the noise of the combined average spectrum will be reduced by a factor of one over the square root of the number of spectra, making any cumulative excess of power due to solar-like oscillations easier to detect against the background power. At the frequencies of interest, the background will be largely dominated by shot noise, but also have a small contribution from granulation. This is what underlies the approach of combining independent spectra, i.e. reducing the variance or standard deviation on the noise. Clearly, the larger the number of stars, the better the prospects of making a detection. However, in practice, targets from a volume-limited field sample will not have the same noise. As we go to greater distances, there will be more stars that can contribute to the combined spectrum, but with higher noise. The combination therefore needs to be a weighted one.

We defined a discrete sample of stars of interest, $i=1... N_{\rm star}$ (see more in Sect.~\ref{sec:pred}). Let $b_i$ be the background power spectral density of a given star across the frequency range where the oscillations are most prominent. The combination of spectra that minimizes the variance of the power spectral density of the ensemble, and hence optimizes the S/N of the combined oscillation signal in the frequency domain, uses weights $1/b_i^2$. The standard deviation of the background power spectral density of the ensemble spectrum is then given by
 \begin{equation}
  \sigma_{\rm PSD} = \left[ \sum_{i=1}^{N_{\rm star}} 1/b^2_i \right]^{-1/2}.
  \label{eq:sigwt}
  \end{equation}
We can then estimate the probability of detecting the combined oscillation signal, given the estimated $\sigma_{\rm PSD}$ and the weighted summation of the oscillation power of the individual targets. Before we go on to explain that step in Sect.~\ref{sec:detect}, we first present the derivation of an analytical approximation of $\sigma_{\rm PSD}$. This analytical form is helpful for understanding the detectability predictions, and can also serve as a useful tool for making quick predictions for other future or planned missions, since it requires only a few basic inputs, as we now go on to see.

\subsection{Analytical approximation for $\sigma_{\rm PSD}$}
\label{sec:predan}

In what follows, we make a number of approximations and idealized assumptions that allow us to derive a straightforward expression for $\sigma_{\rm PSD}$. As noted above, we selected stars over a sufficiently narrow range in $T_{\rm eff}$ on the lower main sequence that we may assume they have similar fundamental properties. For the resulting typical luminosity, $L$, the flux, ${\cal F}$, in the instrumental bandpass will scale in the ideal case with distance, $r$, as 
  \[
  {\cal F} \propto r^{-2}.
  \]
The relative root-mean-square (\textsc{rms}) photometric noise, $\sigma(r)$, scales as
  \[
  \sigma(r) \propto {\cal F}^{-1/2},
  \]
so that
 \[
  \sigma(r) \propto r.
  \]
Consider the noise over a cadence, $\Delta t$. Defining $\sigma_{10}$ to be the resulting photometric noise measured in the instrument bandpass for a target at a distance of 10\,pc having the given $L$ and $T_{\rm eff}$, we may then write
  \begin{equation}
  \sigma(r) = \sigma_{10} \left( \frac{r}{10} \right).
  \label{eq:sig}
  \end{equation}
The background power spectral density, $b(r)$, in the frequency domain for a spectrum satisfying Parseval's theorem is then given by
  \begin{equation}
  b(r)= 2\sigma^2(r) \Delta t \equiv 2\sigma^2_{10} \left(\frac{r}{10}\right)^2 \Delta t,
  \label{eq:ban}
  \end{equation}
In the rest of the analytical treatment in this section, we assume that across the range of frequencies where the oscillation signal is most prominent the contribution from granulation to the background power constitutes only a small fraction of the total background, with shot noise being the dominant source. This is a good approximation for our PLATO targets of interest (see further discussion in Sect.~\ref{sec:detect}), though note that we do include granulation explicitly in the discrete target sample predictions made using the PIC.

Next, we take the number density of stars in the local solar neighbourhood that lie in the targeted parameter range to be $n$ (per $\rm pc^3$), and adopt a fractional sky coverage, $\alpha$, for our observations. Assuming a uniform spatial distribution of targets, the number, $N_{\rm star}(r)$, within a shell of thickness $dr$ at a distance, $r$, will be
  \begin{equation}
  N_{\rm star}(r) = 4\pi \alpha n r^2 dr.
  \label{eq:N}
  \end{equation}
If we construct the weighted sum of the power spectra of all stars from $r=d_{\rm min}$ out to some distance, $r=d_{\rm max}$, the standard deviation of the background power spectral density of the combination will be
 \begin{equation}
  \sigma_{\rm PSD} = \left[ \sum_{d_{\rm min}}^{d_{\rm max}} N_{\rm star}(r)/b^2(r) \right]^{-1/2}.
  \end{equation}
Substituting from Eqs.~\ref{eq:ban} and~\ref{eq:N} above gives
 \begin{equation}
  \sigma_{\rm PSD} = \left[ \left(\ \frac{10}{\sigma_{10}}\right)^{4} \left(\frac{\pi \alpha n}{\Delta t^2}\right) \sum_{d_{\rm min}}^{d_{\rm max}} r^{-2} dr \right]^{-1/2},
  \end{equation}
which in the limit of $dr \rightarrow 0$ implies
  \begin{equation}
  \sigma_{\rm PSD} \simeq \left[ \left(\frac{10}{\sigma_{10}}\right)^{4} \left(\frac{\pi \alpha n}{\Delta t^2}\right) \int\displaylimits_{d_{\rm min}}^{d_{\rm max}} r^{-2} dr \right]^{-1/2}.
  \end{equation}
Evaluating the integral gives 
 \begin{equation}
  \sigma_{\rm PSD} \simeq \left(\frac{\sigma_{10}}{10}\right)^2 \left( \frac{\Delta t}{\sqrt{\pi \alpha n}} \right) \left[ 1/d_{\rm min} - 1/d_{\rm max} \right]^{-1/2}.
  \label{eq:sigwtan}
  \end{equation}
 The form or ‘shape’ of this expression follows intuitive expectations. First, there is the obvious result that a smaller value of $\sigma_{10}$ and higher values of $n$ and $\alpha$ should lead to a reduced standard deviation in the combined spectrum. The dependence on $d_{\rm min}$ and $d_{\rm max}$ also follows expectations. The closest target sets the baseline for the combined spectrum; hence, a smaller $d_{\rm min}$ should give lower noise. Extending to larger distances should also be beneficial, although the one-over-square-root dependence on $d_{\rm max}$ indicates that incremental gains diminish as one incorporates successively more distant stars, reflecting the trade-off at larger distances between having greater numbers of stars to include at the cost of them having higher levels of noise.

\subsection{Detection prediction approach for the ensemble spectrum}
\label{sec:detect}

We followed the now well-established approach (\citealt{Chaplin2011, Campante2016, Lund2016, Schofield2019, Goupil2024}) to predicting the detectability by considering the significance at which the total observed power due to the oscillations, $P_{\rm tot}$, can be measured against background power due to sources of noise and stellar granulation. However, here we must handle the ensemble case and so develop a new variant on the established framework.

The statistics describing the distribution of power within any independent frequency bin of the weighted combination are non-trivial. The combination acts as a $\chi^2$ distribution with an effective number of degrees of freedom, $n_{\rm d.o.f.} \lesssim 30$, i.e. the noise about the underlying combined limit spectrum is not Gaussian but has an asymmetric spread. The effective number of degrees of freedom depends on the respective weights and underlying noise of the constituent spectra that make up the weighted combination. We may simplify the statistics by considering what happens if instead we average or sum the power across a range of bins in the combined, weighted spectrum. Thanks to the central limit theorem, the average or sum will then follow statistics of a normal (Gaussian) distribution about the underlying limit spectrum of the combination. This allows us to develop the following simple prediction for the detectability of the cumulative excess of power due to the oscillations against the broad-band background noise. 

Let us start from the basic case of a single, individual spectrum. Consider a range of frequencies in which the spectrum is dominated by broad-band background noise, having an underlying power spectral density, $\left< b \right>$. The uncertainty on the average power spectral density across the background range, which we assume covers $N$ independent frequency bins, is given by
 \begin{equation}
 \delta_{\left< b \right>} = \frac{\sigma_{\left< b \right>}}{\sqrt{N}} \equiv \frac{\left< b \right>}{\sqrt{N}}.
 \end{equation}
Here, $\sigma_{\left< b \right>}$ is the standard deviation of the noise about the limit spectrum, and as was noted previously, for the $\chi^2$ two-degrees-of-freedom statistics governing the individual spectrum case, $\sigma_{\left< b \right>} = \left< b \right>$.

Next, let $\left< P \right>$ be the average power spectral density across the frequency range, $W$, of the spectrum that contains the oscillation signal. This range is centred on the frequency of the maximum oscillation power, $\nu_{\rm max}$, which follows the scaling relation
 \begin{equation}
 \nu_{\rm max} = \nu_{\rm max,\odot} \left(\frac{M}{M_{\odot}}\right) \left(\frac{R}{R_{\odot}}\right)^{-2}
 \left(\frac{T_{\rm eff}}{T_{\rm eff,\odot}}\right)^{-1/2},
 \label{eq:numax}
 \end{equation}
where we adopt $\nu_{\rm max,\odot} = 3090\,\rm \mu Hz$ and $T_{\rm eff,\odot} = 5777\,\rm K$. Note that the background range -- which is dominated by contributions from shot and instrumental noise, and granulation -- is assumed to lie adjacent to or to straddle the oscillation range. For simplicity, in what follows we assume that the oscillation range also covers $N$ bins, as per the background range. We therefore have
 \begin{equation}
 \delta_{\left< P \right>} \approx \frac{\left< P \right>}{\sqrt{N}} \equiv \frac{\sigma_{\left< P \right>}}{\sqrt{N}}.
 \end{equation}
This relation is approximate: unlike the background case -- in which the underlying power spectral density varies smoothly and slowly in frequency -- this range shows more underlying variation due to the presence of the modes. 

Now, the total excess power due to oscillations in the target range is given by
 \begin{equation}
 P_{\rm tot} = W \left( \left<P\right> - \left< b \right> \right).
 \end{equation}
The uncertainty on the total excess power then follows from simple error propagation, i.e. 
 \begin{equation}
 \delta_{P_{\rm tot}}^2 = W^2 \left( \delta_{\left<P\right>}^2 + \delta_{\left< b \right>}^2 \right).
 \end{equation}
For the main-sequence stars of interest here, the total power due to the oscillations will be a small fraction of the background power due to the PLATO shot and instrumental noise, as was/is the case for the photometric data collected by \emph{Kepler} and TESS (e.g. see \citealt{Chaplin2011, Chaplin2013white, Campante2016, Schofield2019}). This means that $\left< P \right> \simeq \left< b \right>$, allowing us to make the simplification  $\delta_{\left<P\right>} \simeq \delta_{\left< b \right>}$. We may therefore write
\begin{equation}
\delta_{P_{\rm tot}} \simeq \sqrt{2} W \delta_{\left< b \right>} \equiv \sqrt{2} W 
                               \frac{\sigma_{\left< b \right>}}{\sqrt{N}}
                               \equiv \sqrt{2} W
                               \frac{\left< b \right>}{\sqrt{N}}.
\end{equation}
Now, the frequency range, $W$, covered by the $N$ independent frequency bins is given by
 \begin{equation}
 W = N \Delta_T \equiv \frac{N}{T}.
 \end{equation}
Substituting from above then gives
\begin{equation}
\delta_{P_{\rm tot}} \simeq \sigma_{\left< b \right>} \left( \frac{2W}{T}\right)^{1/2} 
                               \equiv \left< b \right> \left( \frac{2W}{T}\right)^{1/2}.
\end{equation}
The implied statistical significance of the excess $P_{\rm tot}$ is therefore
\begin{equation}
 s_{\rm tot} = \frac{P_{\rm tot}}{\delta_{P_{\rm tot}}} 
                    \simeq \left( \frac{P_{\rm tot}}{\sigma_{\left< b \right>}} \right) \left( \frac{T}{2W} \right)^{1/2}
                    \equiv \left( \frac{P_{\rm tot}}{\left< b \right>} \right) \left( \frac{T}{2W} \right)^{1/2}.
 \label{eq:errbasic}
 \end{equation}
Let us now extend to the ensemble case of the weighted combined spectrum. Equation~\ref{eq:sigwt} describes the standard deviation of the background power spectral density, $\sigma_{\rm PSD}$, of the weighted combination. We therefore just need to substitute $\sigma_{\rm PSD}$ for $\sigma_{\left< b \right>}$ in Eq.~\ref{eq:errbasic} above to give the implied statistical significance, $s_{\rm PSD}$, of the excess $P_{\rm tot}$ in the combined, weighted case. The result is
 \begin{equation}
 s_{\rm PSD} \simeq \left( \frac{P_{\rm tot,PSD}}{\sigma_{\rm PSD}} \right)
                           \left( \frac{T}{2W} \right)^{1/2},  
 \label{eq:spsd}
 \end{equation}
where
\begin{equation}
  P_{\rm tot,PSD} = \frac{ \left[ \sum_{i=1}^{N_{\rm star}} P_{\rm tot,i}/b^2_i \right]}{\left[ \sum_{i=1}^{N_{\rm star}} 1/b^2_i \right]}
  \label{eq:ptotwt}
  \end{equation}
is the weighted combination of each individual $P_{\rm tot,i}$. If instead we adopt the analytical approximation of $s_{\rm PSD}$ given by Eq.~\ref{eq:sigwtan}, we obtain
 \begin{equation}
 s_{\rm PSD} \approx  P_{\rm tot} \left(\frac{100}{\sigma_{10}^2\Delta t}\right) \left( \frac{\pi \alpha n T}{2W} \right)^{1/2} \left[ 1/d_{\rm min} - 1/d_{\rm max} \right]^{1/2},
 \label{eq:spsdan}
 \end{equation}
where $P_{\rm tot}$ is now computed from the ‘typical’  or central adopted fundamental properties (see Sect.~\ref{sec:predan}).

As a final step, we can convert $s_{\rm PSD}$ into an equivalent detection probability, $p_{\rm final}$. For a given one-sided false-alarm probability, $p_{\rm false}$, the corresponding critical significance level is 
\begin{equation}
z_{\rm crit} = \Phi^{-1}(1 - p_{\rm false}),
\end{equation}
where $\Phi$ denotes the cumulative distribution function (CDF) of the standard normal distribution. The probability that the measured signal exceeds this critical threshold is then
\begin{equation}
p_{\rm final} = 1 - \Phi\!\left(z_{\rm crit} - s_{\rm PSD}\right).
\label{eq:pfinal}
\end{equation}

We now go on to explain how we used the formulae above to make ensemble predictions for PLATO.

\section{Predictions for PLATO}
\label{sec:pred}

The PLATO instrument comprises 26 small-aperture (12-cm) wide-field telescopes. The 24 normal cameras provide data on a $\Delta t = 25\,\rm sec$ cadence, with a pixel scale of 15\,arcsec. The normal cameras are grouped into four co-aligned sets, with the pointings slightly offset to give partially overlapping fields and a combined effective field of view (e.g. see \citealt{Nascimbeni2025}) of ${\simeq}2149\,\rm deg^2$. Depending on their placement in the combined field, targets may therefore be observed by 6, 12, 18, or 24 normal cameras. There are in addition two fast cameras, which provide data on a 2.5-sec cadence in blue and red filters. Here, we make predictions assuming observations made by the normal cameras.

PLATO will begin science operations with a $T=2\,\rm yr$ stare in its southern LOPS2 field (\citealt{Montalto2021, Nascimbeni2022, Nascimbeni2025}). Decisions on what follows will be made after launch, but one scenario would see observations of this field extended for a full 8\,yr. Here, we produce predictions for $T=2$, 4, and 8-yr observations of bright K dwarfs in the LOPS2 field, using data and information contained in the PIC\footnote{We used PIC version 2.1.0.1}.

\subsection{Underpinning data}
\label{sec:data}

We used the PIC estimates of $M$, $R$, and $T_{\rm eff}$ to calculate oscillation and granulation parameters for selected targets, and the expected total noise per hour (e.g. see \citealt{Boerner2024}) given in the PIC to calculate the background power spectral density of each target that is attributable to noise. Note that we excluded targets that have a Gaia renormalized unit weight error (RUWE) metric $>3.6$ to remove sources with non-single-star astrometric solutions, thereby rejecting objects likely affected by unresolved multiplicity or strong blending, which could also give unreliable stellar parameter estimates.

The main PLATO predictions come from calculating the discrete, weighted summations for $\sigma_{\rm PSD}$ (Eq.~\ref{eq:sigwt}) and $P_{\rm tot,PSD}$ (Eq.~\ref{eq:ptotwt}), using inputs from the PIC for each individual star in the target sample, from which we can then estimate the detection significance, $s_{\rm PSD}$, and equivalent detection probability, $p_{\rm final}$, of the weighted ensemble using Eqs.~\ref{eq:spsd} and~\ref{eq:pfinal}, respectively. We supplement these main results with predictions made using the analytical approximations (Eqs.~\ref{eq:sigwtan} and~\ref{eq:spsdan}).

\subsection{Predicted parameters for each target}
\label{sec:params}

In this section we explain in detail how we calculated the necessary oscillation, granulation, and background parameters for each individual target, and the effective number of degrees of freedom needed to determine the detection significance and probability.

\subsubsection{Oscillation power}
\label{sec:oscpow}

Following convention (e.g. see \citealt{Chaplin2011, Campante2016, Schofield2019, Goupil2024}), we assumed that the power excess due to the oscillations follows a Gaussian-like envelope. The total oscillations power, $P_{\rm tot}$, can then be written in the form
 \begin{equation}
 P_{\rm tot} = \left( \frac{\pi}{4\ln 2} \right)^{1/2} H_{\rm env} \Gamma_{\rm env},
 \end{equation}
where $H_{\rm env}$ and $\Gamma_{\rm env}$ are, respectively, the height and full width at half maximum (\textsc{FHWM}) of the envelope. The envelope height follows from (e.g. see \citealt{Basu2017})
 \begin{equation}
 H_{\rm env} = \left(\frac{\varsigma A_{\rm max}^2}{\Delta\nu} \right),
 \end{equation}
where $A_{\rm max}$ is the root mean square (\textsc{RMS}) maximum equivalent radial-mode amplitude, $\varsigma$ is the sum in power of the mode visibilities of different angular degrees, and
 \begin{equation}
 \Delta\nu = \Delta\nu_{\odot} \left( \frac{M}{M_{\odot}} \right)^{1/2} \left( \frac{R}{R_{\odot}} \right)^{-3/2}  
 \end{equation}
is the large frequency separation, and we adopted $\Delta\nu_{\odot} = 135\,\rm \mu Hz$. We followed \citet{Chaplin2011} in assuming that the maximum amplitudes follow
 \begin{equation}
 A_{\rm max} = A_{\rm max,\odot} \beta \left( \frac{L/M}{L_{\odot}/M_{\odot}} \right)^s 
                \left(\frac{T_{\rm eff}}{T_{\rm eff,\odot}} \right)^{-2},
 \end{equation}
where $s$ is the aforementioned amplitude exponent, and $\beta$ is a factor that corrects for the overestimation of mode amplitudes in F-type stars. Since here we deal with much cooler stars, we are safe in setting $\beta = 1$.

Both $A_{\rm max}$ and $\varsigma$ are sensitive to the bandpass filter of the observations. Here we used calculations for the PLATO bandpass made by \citet{Lund2026}, which yield values of $A_{\rm max,\odot} = 2.4\,\rm ppm$ and $\varsigma = 3.15$. 

For the envelope width, we adopted the scaling of \citet{Mosser2012}:
 \begin{equation}
 \Gamma_{\rm env} = 0.66\, \numax^{0.88} \equiv 
 0.66\, \left(\frac{M}{M_{\odot}}\right)^{0.88} \left(\frac{R}{R_{\odot}}\right)^{-1.76}
 \left(\frac{T_{\rm eff}}{T_{\rm eff,\odot}}\right)^{-0.44}.
 \end{equation}
Cool main-sequence stars with good asteroseismic detections have been shown to follow this relation to reasonable approximation, and whilst there is clearly an additional temperature dependence to the observed widths not captured by the above, that dependence only appears to be relevant at higher $T_{\rm eff}$ than we consider here (see \citealt{Schofield2019thesis}).

Finally, we note that we took no account in our predictions of the impact of magnetic activity on the mode amplitudes (e.g. see \citealt{Chaplin2011act, Campante2014, Mathur2019, Sayeed2025}). It is worth noting that the amplitude scaling relation used here is calibrated on measurements from \emph{Kepler} targets that span a range of activity levels. We return to this point in Sect.~\ref{sec:conc}.

\subsubsection{Background power}
\label{sec:backpow}

The mean background power spectral density, $\left< b \right>$, has contributions from shot and instrumental noise, $b_{\rm instr}$, and from granulation, $b_{\rm gran}$, i.e. 
 \begin{equation}
 \left< b \right> = b_{\rm instr} + b_{\rm gran}.  
 \end{equation}
The shot and instrumental contribution follows from
  \begin{equation}
  b_{\rm instr} = 2\sigma_{\Delta t}^2 \Delta t,
  \label{eq:binstr}
  \end{equation}
where $\sigma_{\Delta t}$ is the noise per cadence, $\Delta t$. Here, we assumed observations would in practice be made at a PLATO cadence of 25\,sec. This rapid cadence -- which results in a very high Nyquist frequency of $20,000\,\rm \mu Hz$ -- allows us to ignore the attenuation of oscillation signals at high frequencies due to the finite sampling in the time domain (e.g. see \citealt{Basu2017} and references therein). 

As is standard for PLATO, the PIC provides noise estimates for targets on a 1-hr timescale -- values that we refer to here as $\sigma_{\rm hr}$ -- which capture all expected sources of instrumental noise. Using these 1-hr values, the above was modified to
  \begin{equation}
  b_{\rm instr} = 2\sigma_{\rm hr}^2 \times 3600 \times 10^{-6},
  \label{eq:binstr1}
  \end{equation}
where the factor $10^{-6}$ ensures the estimates are in units of $\rm ppm^2\,\mu Hz^{-1}$.

We followed \citet{Chaplin2011} when modelling the granulation contribution. Here, our term $b_{\rm gran}$ corresponds to the power spectral density of the granulation at $\nu_{\rm max}$. As such, we ignored the frequency dependence of the granulation signal, taking this as a suitable proxy value. It follows from the relation
 \[
 b_{\rm gran} = b_{\rm gran,\odot} \left( \frac{\nu_{\rm max}}{\nu_{\rm max,\odot}} \right)^{-2} 
 \]
 \begin{equation}
 ~~~~~~~~~\equiv
 b_{\rm gran,\odot}\left(\frac{M}{M_{\odot}}\right)^{-2} \left(\frac{R}{R_{\odot}}\right)^{4}
  \left(\frac{T_{\rm eff}}{T_{\rm eff,\odot}}\right).
 \end{equation}
Note that $b_{\rm gran,\odot} = 0.09\,\rm ppm^2 \mu Hz^{-1}$ in the PLATO bandpass \citep[cf.][]{Lund2026}.

\subsubsection{Degrees of freedom for detectability calculation}

Finally, we needed to fix the width, $W$, adopted in computations of the detection significance, $s_{\rm PSD}$, and the associated detection probability, $p_{\rm final}$. This in effect defines the number of degrees of freedom used in the calculations, through the corresponding number of bins, $N$ (Eq.~\ref{eq:N}). A width of $W \simeq 2 \Gamma_{\rm env}$ would seem to be a straightforward, natural choice for any individual star, since this will capture all but a small fraction of the total oscillation power contained within its Gaussian envelope. However, we must bear in mind that the composite spectrum is a weighted summation over stars that have slightly different $\nu_{\rm max}$. This has the effect of spreading power beyond the $2W$ range of any single star. For every combination of stars we considered, we therefore calculated the effective full width of the composite power envelope. We modelled the envelope of each star as a Gaussian -- with $H_{\rm env}$ and $\Gamma_{\rm env}$ fixed accordingly -- then made a weighted combination to produce the composite envelope, from which the width, $W$, was calculated. This is the width we adopted in the predictions that follow.


\begin{figure}
\centering
\centerline{\includegraphics[width=0.5\textwidth]{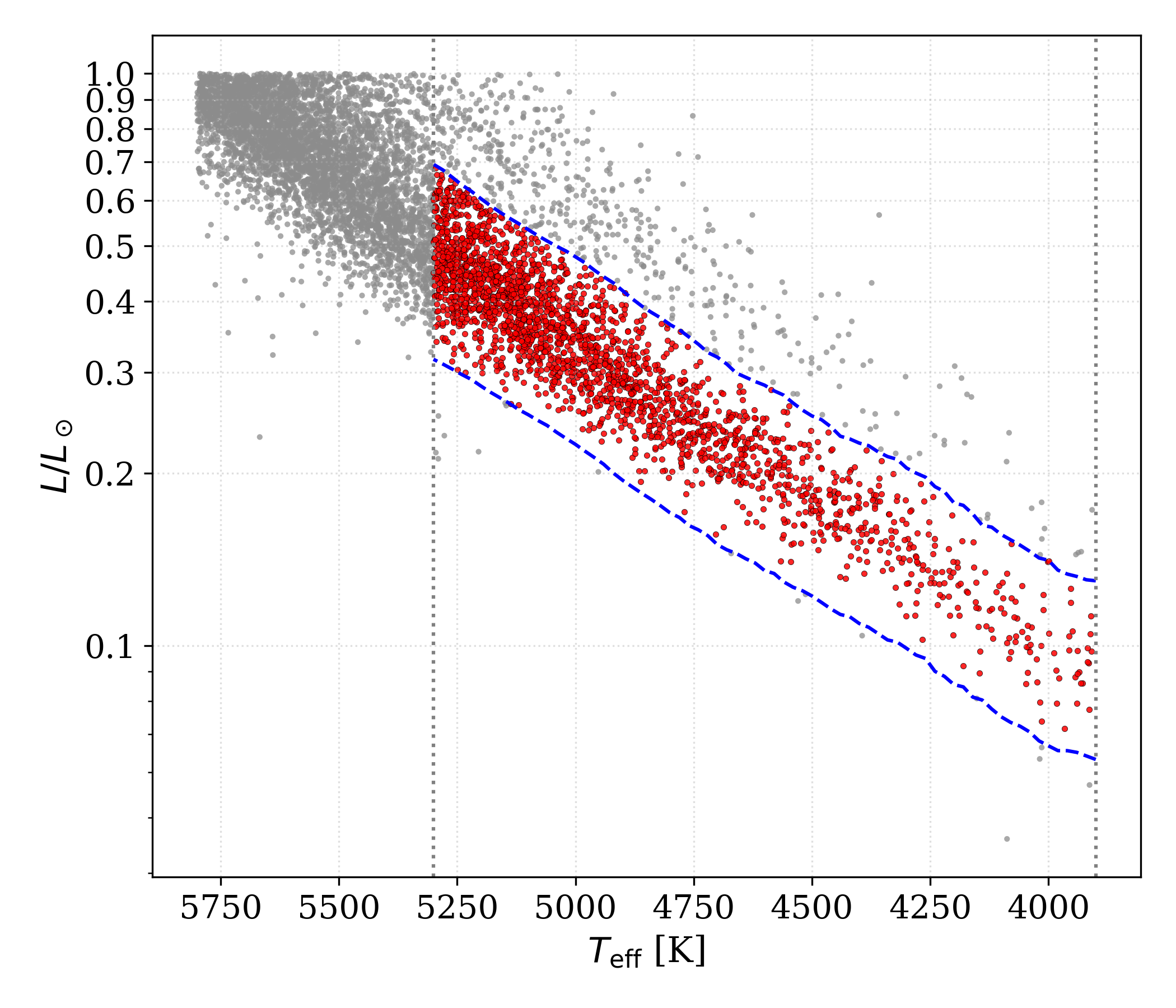}}
        \caption{HR diagram showing stars in the PIC on the lower main sequence that have $V \le 12$. The selected sample of K dwarfs is plotted in red. The dashed blue lines mark thresholds used to reject targets that lie significantly above or below the main sequence (see text).}
          \label{fig:ksample1}
\end{figure}


\subsection{Results}
\label{sec:res}


\begin{figure*}
\centering
\centerline{\includegraphics[width=1.00\textwidth]{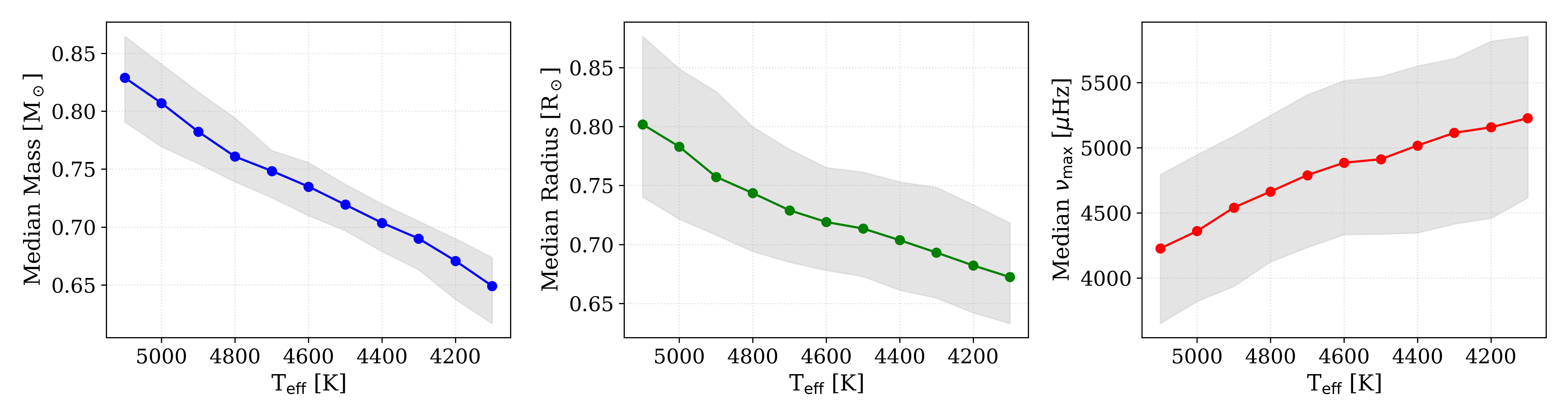}}
\centerline{\includegraphics[width=0.66\textwidth]{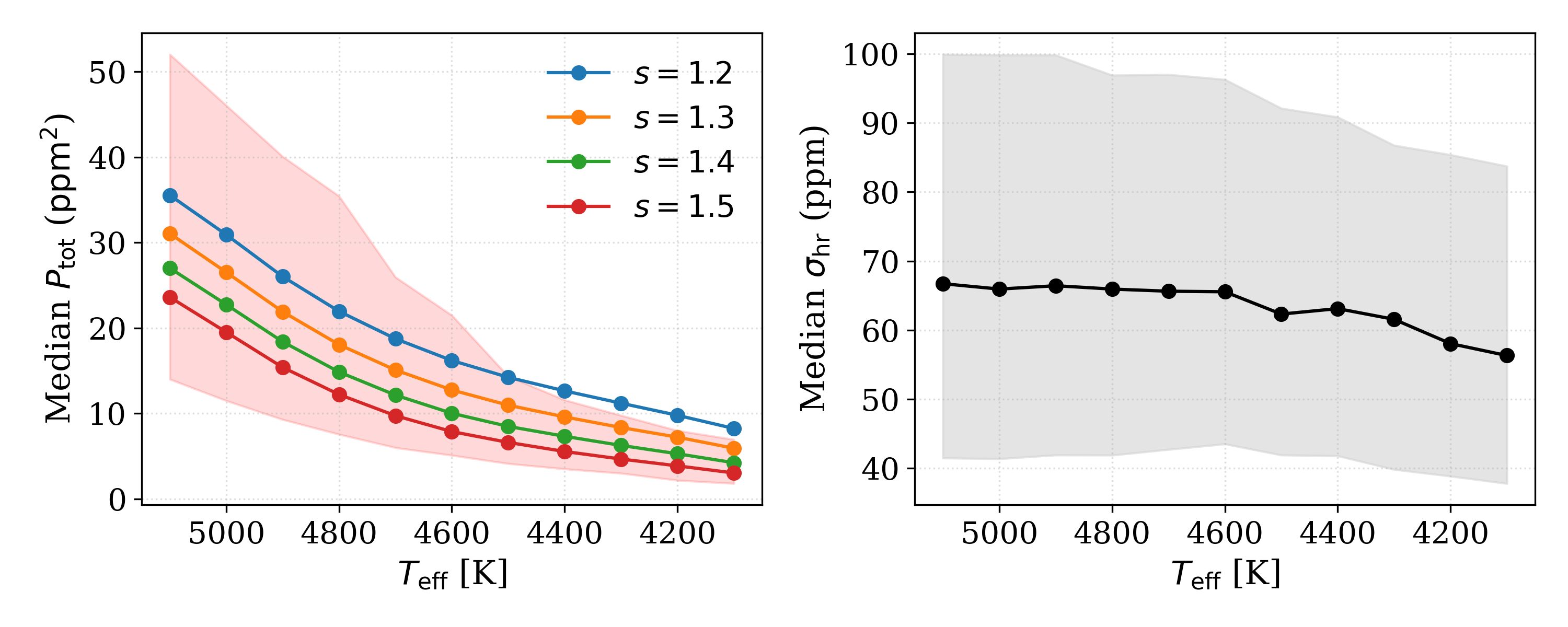}}
        \caption{Median $M$, $R$, $\nu_{\rm max}$ (top row), $P_{\rm tot}$, and $\sigma_{\rm hr}$ (bottom row) of each PIC subsample as a function of the central temperature of the bin. The shaded regions bound the 16th and 84th percentiles. For the $P_{\rm tot}$ plot, the shading is associated with the $s=1.5$ predictions.}
          \label{fig:ksample2}
\end{figure*}


Figure~\ref{fig:ksample1} is an HR diagram showing stars in the PIC on the lower main sequence that have apparent magnitudes in the visual band $V \le 12$. To filter out targets lying significantly above and below the main sequence, we applied a moving median filter of width 400\,K to constrain the underlying $L$-$T_{\rm eff}$ trend, and then offset that trend line by 0.1\,dex and $-0.2\,\rm dex$ to give, respectively, the upper and lower selection thresholds marked by the dashed blue lines in the figure. The final K-dwarf sample comprises 2,410 stars spanning the range $3900 \le T_{\rm eff} < 5300\,\rm K$ -- i.e. from the K to M dwarf boundary at the cool end to the G to K dwarf boundary at the hot end -- with an apparent visual magnitude cut-off of $V \le 12$. In what follows, we make detectability predictions for subsamples of stars in 400\,K wide $T_{\rm eff}$ bins, with each bin offset by 100\,K. The narrower the width of the bins, the more similar the fundamental properties and the $\nu_{\rm max}$ of the targets, which maximizes the extent to which a peaked accumulation of power builds in the combined spectrum when the individual spectra are combined. However, narrowing the bin width comes at the expense of reducing the number of targets whose frequency power spectra will be combined in that $T_{\rm eff}$ bin, which in turn reduces the S/N in the composite spectrum. The choice of 400\,K was found to strike a good compromise between the spread on fundamental properties in each bin and the size of the sample.

Figure~\ref{fig:ksample2} plots the median $M$, $R$, $\nu_{\rm max}$ (top row), $P_{\rm tot}$, and $\sigma_{\rm hr}$ (bottom row) of each subsample as a function of the central temperature of the bin, using the PIC parameters and with the shaded regions bounding the 16th and 84th percentiles. Note that we also tested the impact on our predictions of instead using the \emph{Gaia} DR3 \texttt{gspphot} \citep{Gaia2023} estimates of $\log{g}$, $R$, and $T_{\rm eff}$ for our selected targets. This did not significantly affect the results that follow, or the conclusions drawn from our analysis. 


\begin{figure*}
\centering
\centerline{\includegraphics[width=1.0\textwidth]{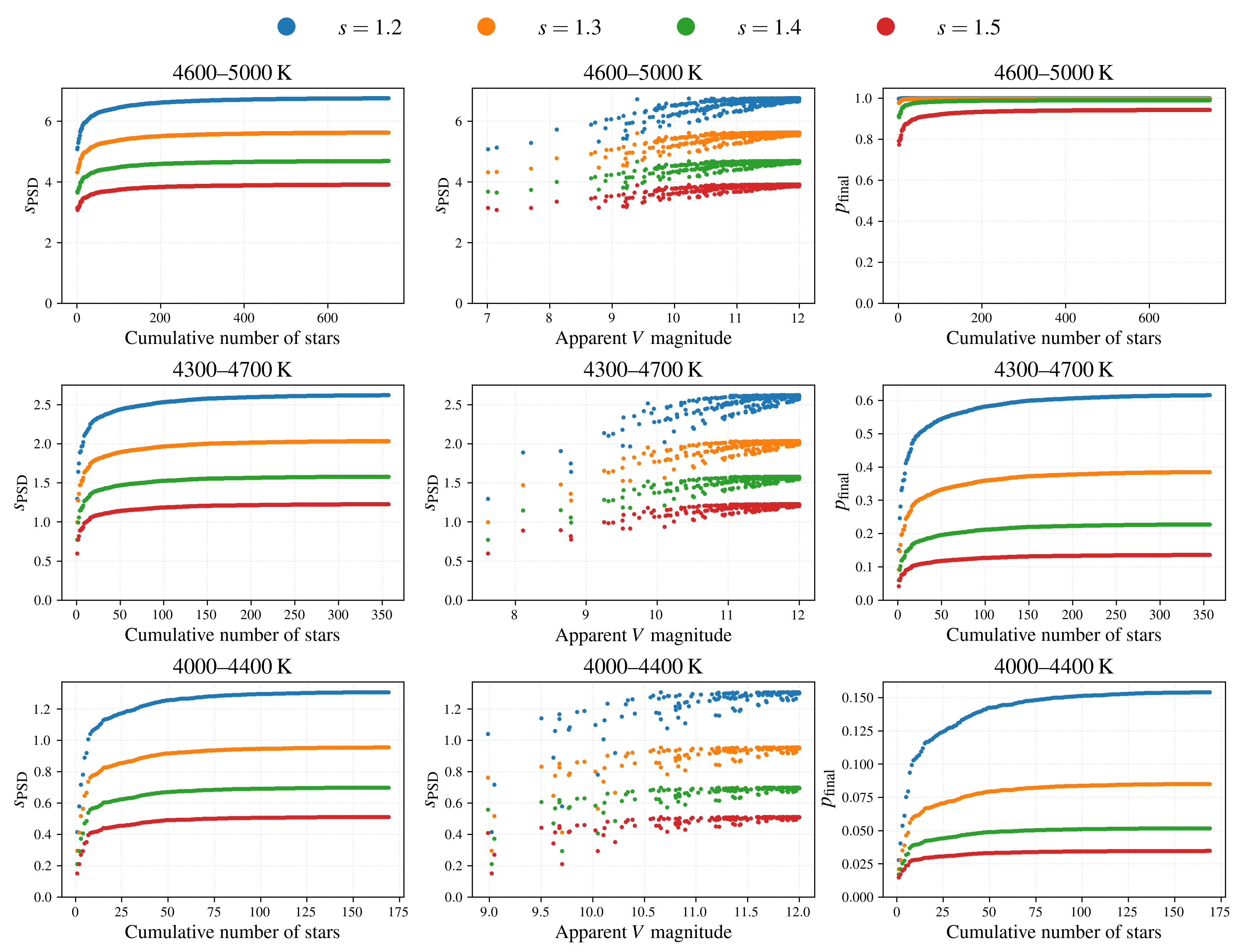}}
        \caption{Prediction results using data from the PIC, for three temperature bins (by row); different values of the amplitude exponent, $s$ (see plot annotation); and an assumed observation duration of $T=2\,\rm yr$. Plots in the left-hand column show $s_{\rm PSD}$ as a function of the cumulative star count, while those in the middle column instead use the apparent visual magnitude, $V$, as the independent variable. Plots in the right-hand column show $p_{\rm final}$ as a function of the cumulative star count.}
          \label{fig:sandp}
\end{figure*}


To introduce our predictions and the general form they take, Fig.~\ref{fig:sandp} shows illustrative results for three temperature bins, different values of the amplitude exponent $s$, and an assumed observation duration of $T=2\,\rm yr$. Plots in the left-hand column show the predicted significance, $s_{\rm PSD}$, as a function of the cumulative number of stars included in the combined spectrum. Stars are added in order of increasing noise, $\sigma_{\rm hr}$, beginning with the lowest-noise target. The middle column shows the predictions as a function of apparent visual magnitude, $V$. The branches in the plots arise from differences in the number of PLATO cameras that observe each target, depending on where in the PLATO field of view the target lies. This means that, for PLATO at least, there is not a one-to-one relationship between noise and apparent magnitude. Plots in the right-hand column show the corresponding detection probabilities, $p_{\rm final}$, calculated for a false-alarm threshold of $p_{\rm false} = 0.01$. 

Results for all three temperature bins in Fig.~\ref{fig:sandp} show similar trends. The detection significance and probability both increase as we include more stars. Most of the cumulative gain comes from the early batches of stars. This is to be expected because they have the lowest noise. The results asymptote at higher star counts: as was noted already near the end of Sect.~\ref{sec:predan}, this reflects the diminishing returns of including higher-noise stars. The significance and probability decrease for higher values of the amplitude exponent, $s$, and as we progress from the highest to the lowest temperature bins, owing to the lower predicted amplitudes.

It is also instructive to show the predictions from the analytical approach. Recall that it uses Eqs.~\ref{eq:sigwtan} and~\ref{eq:spsdan} and requires its own set of inputs. First, the effective PLATO field of view of ${\simeq}2149\,\rm deg^2$ corresponds to an all-sky fraction of $\alpha = 0.052$. Appendix~\ref{sec:num} presents the analysis we performed to arrive at an equivalent stellar number density, $n=2.4 \times 10^{-3}\,\rm pc^{-3}$, for the selected sample in the $4300 \le T_{\rm eff} < 4700\,\rm K$ temperature bin. We then adopted central fundamental stellar properties for the bin of $M=0.72\,\rm M_{\odot}$, $R=0.72\,\rm R_{\odot}$, and $T_{\rm eff} = 4500\,\rm K$. To estimate the noise $\sigma_{10}$ for these central properties, we converted the central luminosity ($L \propto R^2 T_{\rm eff}^4$) to a $V$-band magnitude at 10\,pc, using a bolometric correction calculated following \citet{Torres2010}, and then used the noise model in \citet{Boerner2024} to estimate $\sigma_{\rm hr}$. We made predictions for different numbers of cameras, assuming that the noise scales with the square root of that number.


\begin{figure*}
\centering
\centerline{\includegraphics[width=1.0\textwidth]{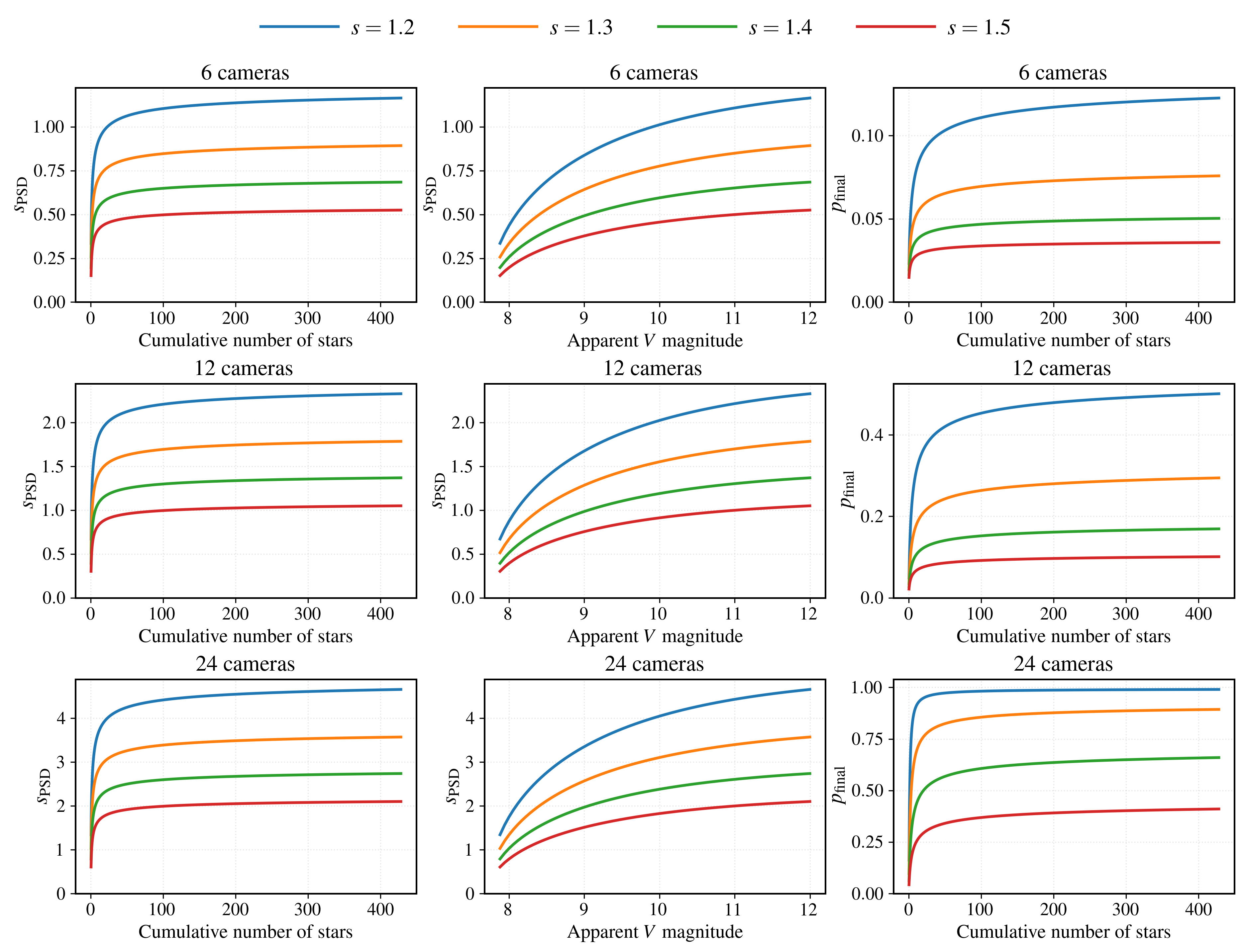}}
        \caption{Estimated $s_{\rm PSD}$ and $p_{\rm final}$ from the analytical predictions for the $4300 \le T_{\rm eff} < 4700\,\rm K$ temperature bin, with different rows showing predictions for different numbers of PLATO cameras.}
          \label{fig:4347anal}
\end{figure*}


Figure~\ref{fig:4347anal} plots the analytical predictions of $s_{\rm PSD}$ and $p_{\rm final}$ for the $4300 \le T_{\rm eff} < 4700\,\rm K$ temperature bin, with different rows showing predictions for different numbers of PLATO cameras. The form of the predictions follows what we saw from the discrete numerical predictions. Note the cumulative count of stars reaches a slightly higher number at $V=12$ than the PIC subsample for this temperature bin, because the subsample is consistent with being volume complete only out to a distance of ${\simeq} 87\,\rm pc$ (see Appendix~\ref{sec:num}). However, a star having the adopted central fundamental parameters used to make the analytical predictions in Fig.~\ref{fig:4347anal} reaches $V=12$ at a slightly greater distance of just over 93\,pc. 

As expected, the results from the analytical predictions are highly dependent on the number of cameras. The 12-camera prediction is the closest to the discrete numerical predictions in Fig~\ref{fig:sandp}. This is not surprising: 12 corresponds to the median number of observing cameras (out of a total of 24) expected over any PLATO field sample\footnote{Of the total PLATO field of view, 13.5\,\% has 24-camera coverage, 11.1\,\% has 18-camera coverage, 32.9\,\% has 12-camera coverage, and 42.5\,\% has 6-camera coverage. Strictly speaking there is therefore a 57.5\,\% chance of having coverage by 12 or more cameras.}. However, the 12-camera predictions will err on the slightly pessimistic side should the lowest-noise star be observed with a larger number of cameras, i.e. that star provides the anchor point from which the combined $s_{\rm PSD}$ accumulates.


\begin{figure*}
\centering
\centerline{\includegraphics[width=0.85\textwidth]{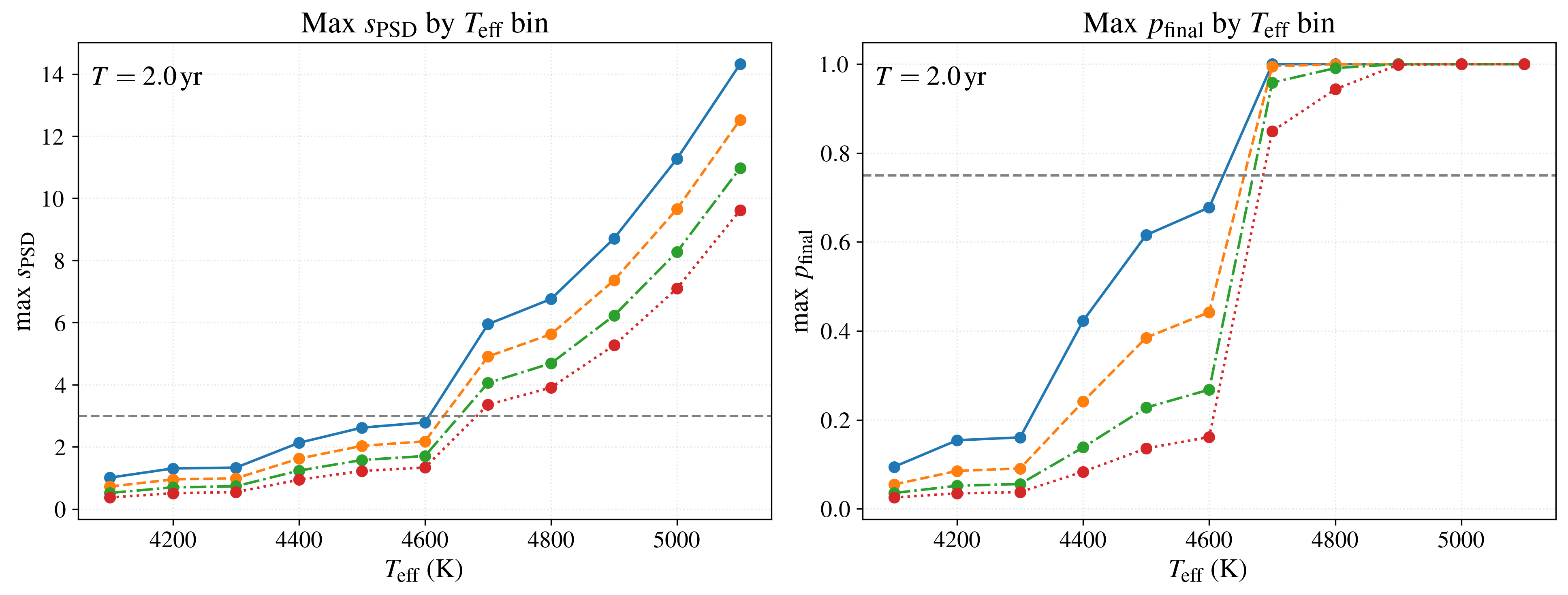}}
\centerline{\includegraphics[width=0.85\textwidth]{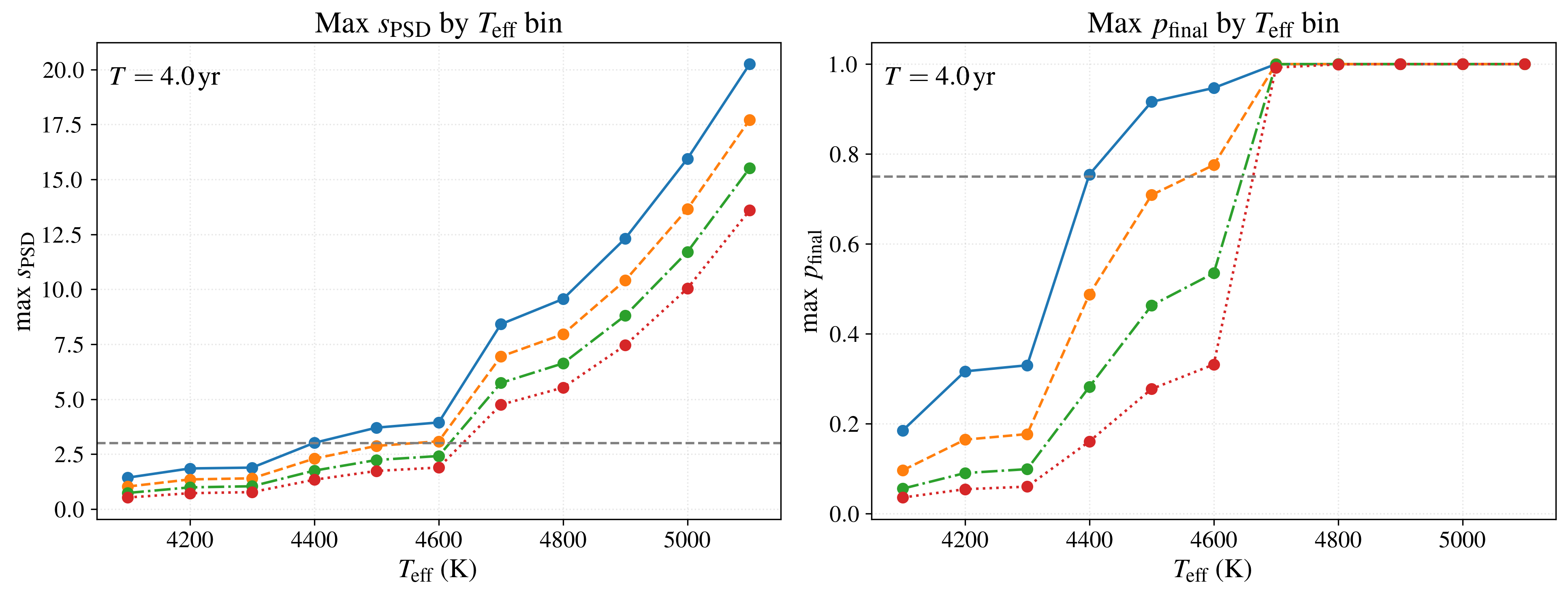}}
\centerline{\includegraphics[width=0.85\textwidth]{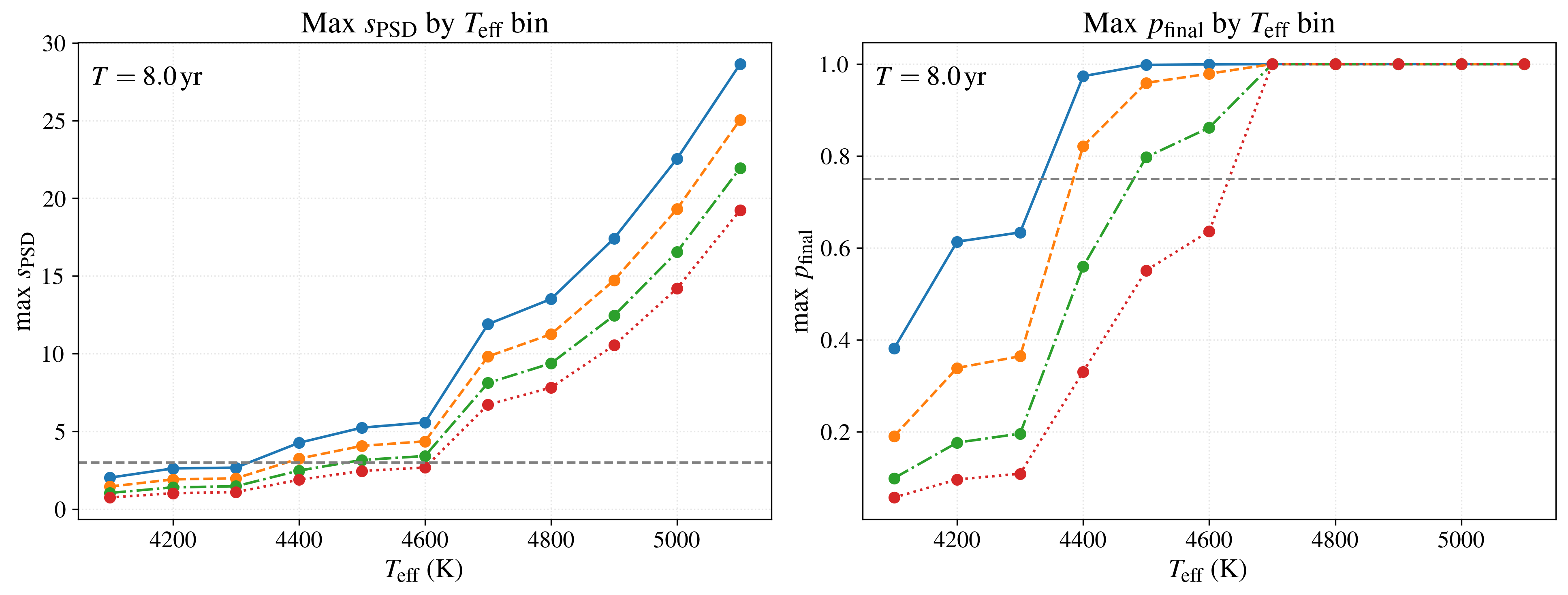}}
        \caption{Maximum $s_{\rm PSD}$ (left panels) and $p_{\rm final}$ (right panels) to which the predictions asymptote as more stars are included in the combination, plotted as a function of the central temperature of each bin. Predictions are shown for observations of length $T=2$, 4, and 8\,yr. Colours denote different values of the amplitude exponent, $s$, and are as per previous figures, i.e. $s=1.2$ in blue, $s=1.3$ in orange, $s=1.4$ in green, and $s=1.5$ in red.}
          \label{fig:final}
\end{figure*}


Figure~\ref{fig:final} captures the bottom-line predictions for PLATO. Here, we plot the maximum $s_{\rm PSD}$ (left-hand panels) and $p_{\rm final}$ (right-hand panels) to which the predictions asymptote as more stars are added, as a function of the central temperature of each bin. We show predictions for observations of length $T=2$, 4, and 8\,yr (top to bottom rows). The horizontal dashed lines shown in the left-hand panels mark the $3\sigma$ level. The horizontal lines in the right-hand panels mark the corresponding probability, $p_{\rm final}$, that an underlying $3\sigma$ signal will beat the adopted false alarm threshold of $p_{\rm false} = 0.01$.

With 2\,yr of data, the detection significance reaches the $3\sigma$ level for all values of the amplitude exponent when the central bin temperature $T_{\rm eff} \gtrsim 4700\,\rm K$. That the trends are not entirely smooth is just a reflection of the finite size and nature of the underlying samples in each temperature bin. With 4\,yr of data, what were marginal detections become firm, highly significant detections, and the $s=1.4$ data now exceed the $3\sigma$ level at a central bin temperature of $T_{\rm eff} \gtrsim 4600\,\rm K$. With even more data we predict significant detections at $s=1.4$ and below that push into the late-K regime. 

\begin{figure*}
\centering
\centerline{\includegraphics[width=0.45\textwidth]{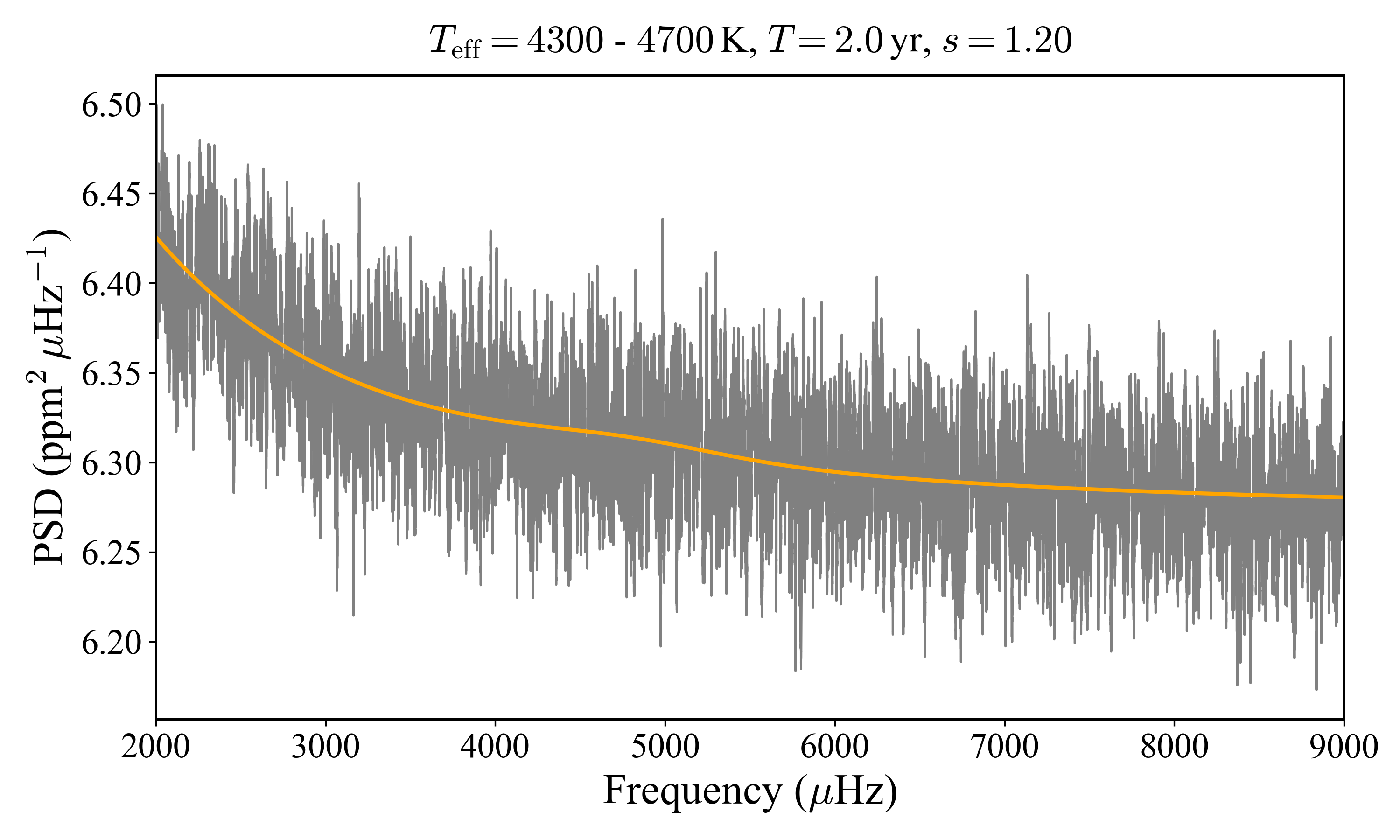}
            \includegraphics[width=0.45\textwidth]{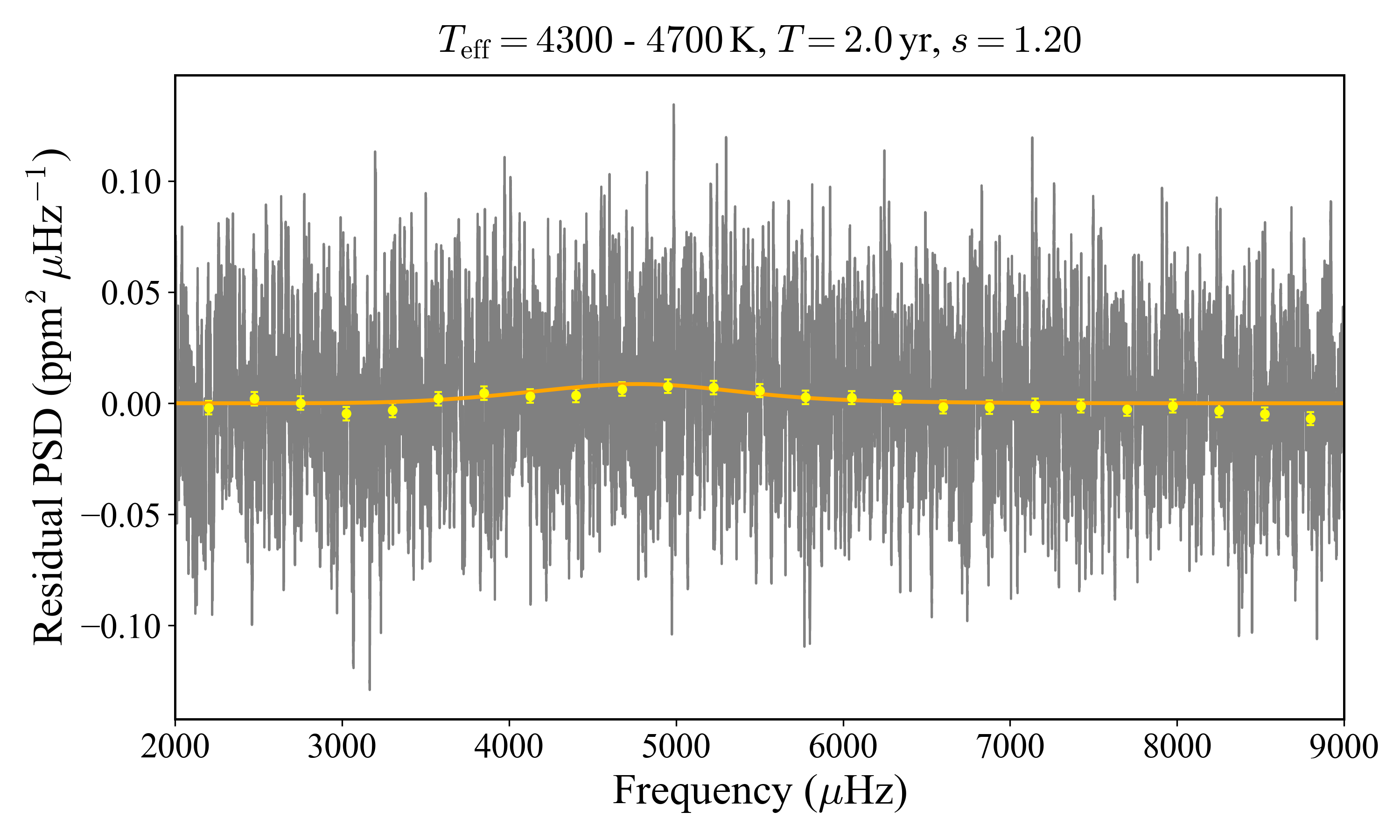}}
\centerline{\includegraphics[width=0.45\textwidth]{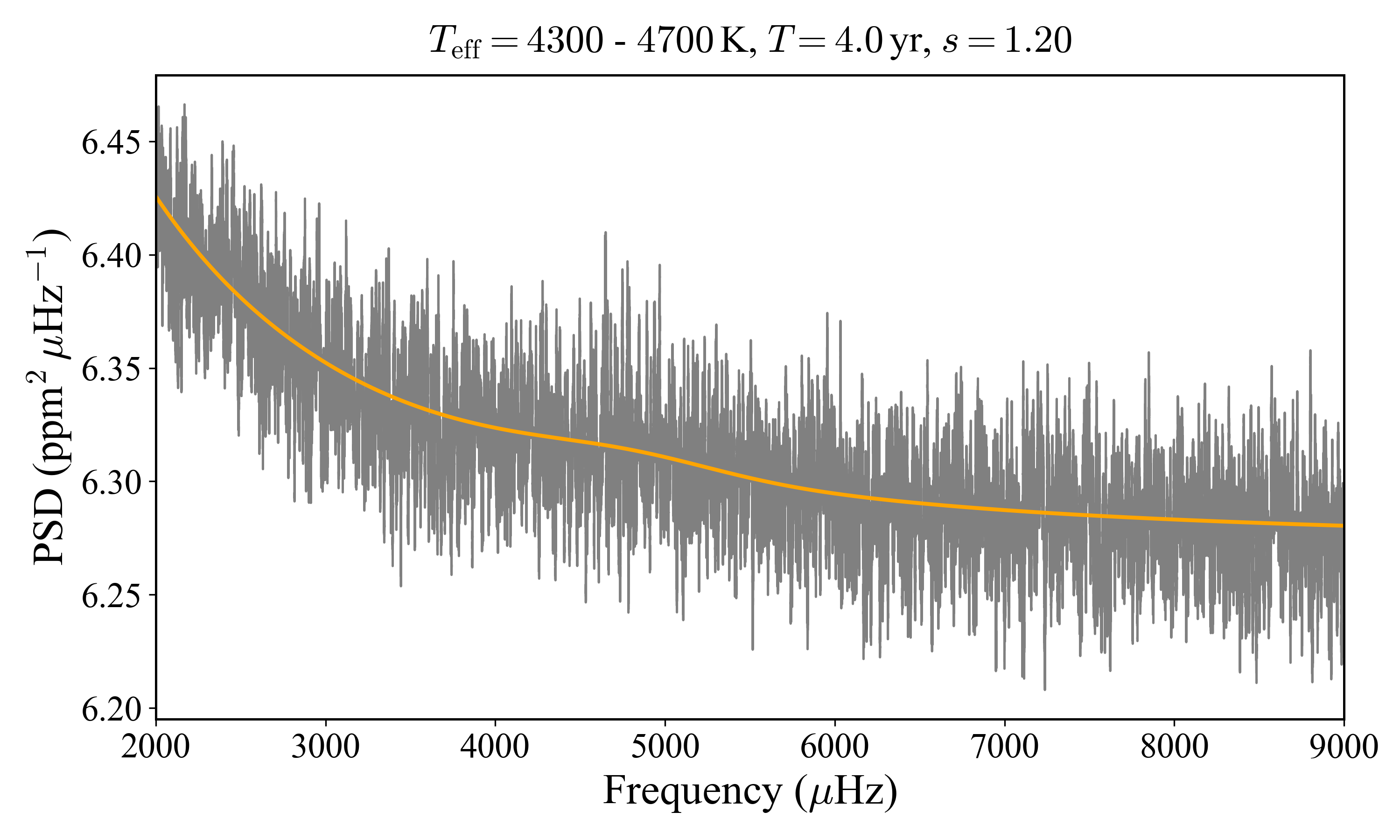}
            \includegraphics[width=0.45\textwidth]{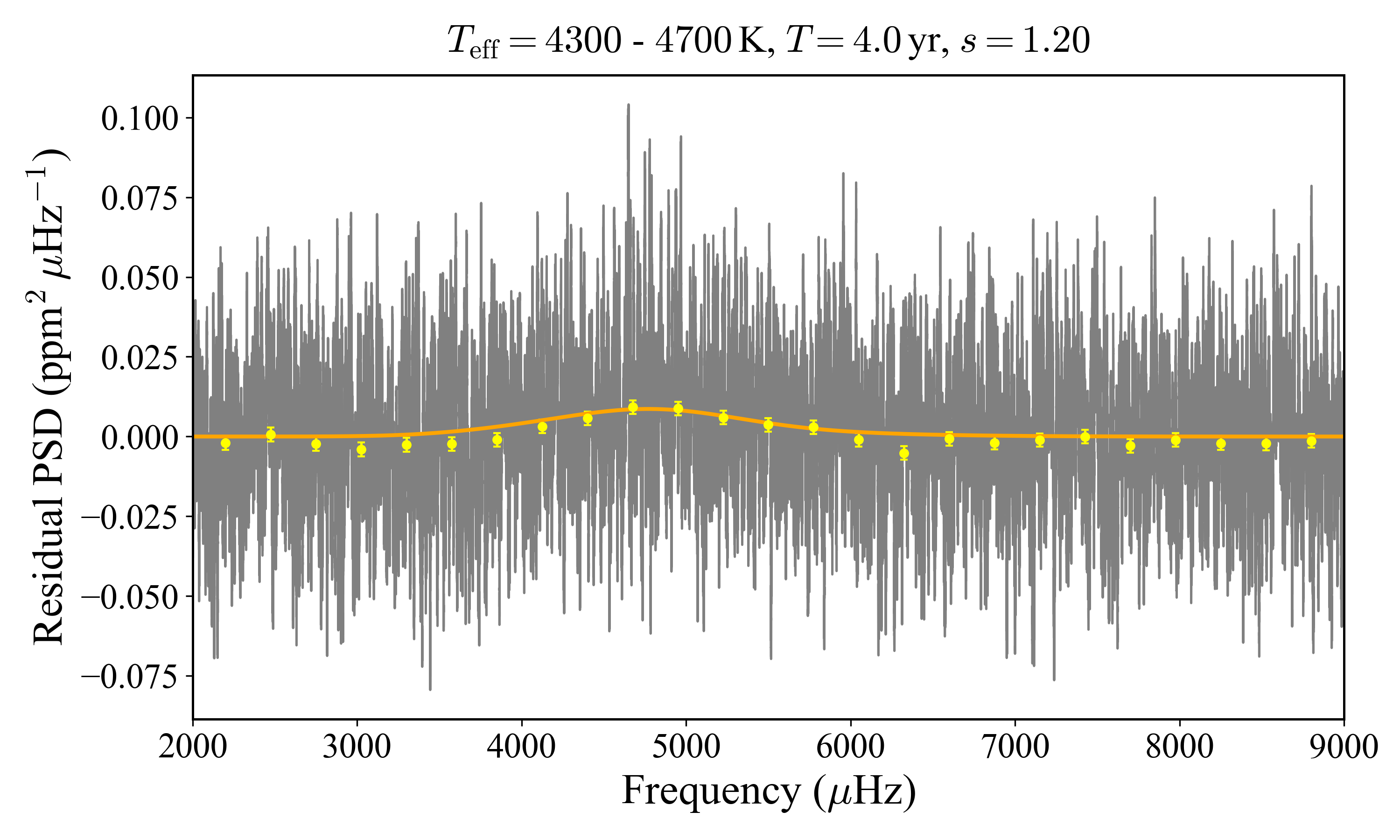}}
        \caption{Artificial PLATO spectra for the $4300 \le T_{\rm eff} < 4700\,\rm K$ temperature bin, calculated assuming $s=1.2$ and observation lengths of $T=2\,\rm yr$ (top panels) and $T=4\,\rm yr$ (bottom panels). Left panels: Combined, weighted spectra for all 357 stars in this bin (rendered in grey), after applying a $5\,\rm \mu Hz$ boxcar to smooth the spectrum. The orange lines correspond to the underlying combined limit spectrum, with the p-mode contribution corresponding to the weighted combination of the Gaussian envelopes of each star. Right panels: Residuals given by subtracting the background limit spectra from the combined spectra. This should leave the weighted, combined contribution due to the p modes, whose underlying limit contribution is shown by the orange line. The yellow points with error bars correspond to averages of the combined spectra computed for $550\,\rm \mu Hz$-wide bins, each offset by $275\,\rm \mu Hz$.}
          \label{fig:artspec}
\end{figure*}

Figure~\ref{fig:artspec} shows examples of what combined the power spectra would look like. Here, we constructed artificial PLATO spectra for the $4300 \le T_{\rm eff} < 4700\,\rm K$ temperature bin, including contributions from p-modes, granulation, and noise. Detail on the construction is presented in Appendix~\ref{sec:artspec}. Artificial spectra were calculated assuming $s=1.2$, and observation lengths of $T=2\,\rm yr$ (top panels), corresponding to a very marginal detection case (see Fig.~\ref{fig:final}), and $T=4\,\rm yr$ (bottom panels), corresponding to a firmer detection. The left-hand panels show the combined, weighted spectra for all 357 stars in this bin (rendered in grey), after applying a $5\,\rm \mu Hz$ boxcar to smooth the spectrum.  The orange lines correspond to the underlying combined limit spectrum, with the p-mode contribution corresponding to the weighted combination of the Gaussian envelopes of each star (see Sect.~\ref{sec:oscpow}). The right-hand panels plot the residuals given by subtracting the background limit spectra from the combined spectra. This should leave the weighted, combined contribution due to the p modes, whose underlying limit contribution is shown by the orange line. The yellow points with error bars correspond to averages of the combined spectra computed for $550\,\rm \mu Hz$-wide bins, each offset by $275\,\rm \mu Hz$.


\section{Conclusions}
\label{sec:conc}

We have presented an ensemble method of detecting signatures of solar-like oscillations in K-type dwarfs, and shown that the upcoming PLATO mission provides an ideal opportunity to exploit this new approach. Frequency power spectra on hundreds of K dwarfs lying in constrained ranges of effective temperature would be combined in a weighted manner to significantly improve the detectability of the oscillations. Whilst this approach means one is unable to extract usable constraints on individual oscillation frequencies, it provides a way to detect and measure the characteristics of the composite envelope of oscillation power given by the ensemble.

Our predictions -- which use information from the PIC on bright K-dwarf targets in the southern LOPS2 field -- suggest that PLATO has the potential to provide ensemble detections well into the K-dwarf regime. The effective temperature limit at which we start to see robust detections would provide valuable information. But it is also worth stressing that even the absence of significant detections would allow us to place strong limits on how the mode amplitudes scale with stellar properties. Results on the stars with known detections suggest that there is a breakpoint at the boundary between G- and K-type dwarfs in how the amplitudes scale with fundamental stellar properties. Our predictions indicate we would be able to explore that transition in some detail using ensemble PLATO data based on a statistically significant sample of stars.  We know that strong magnetic activity suppresses the amplitudes of solar-like oscillations (e.g. see \citealt{Chaplin2011act} and references therein). \citet{Sayeed2025} have recently demonstrated a significant correlation of mode amplitudes with direct measures of activity based on emission in Ca H\&K lines. Activity is, of course, likely to be a relevant issue for any field sample of K dwarfs. However, a large statistical sample allows the effects to be explored; for example, by breaking down the large samples into subsamples that show different levels of activity based on additional spectroscopic data or signatures of rotational modulation of active features in the PLATO light curves. 

Results from PLATO will of course depend on the quality of the photometry, including the potential impact of background sources. Our predictions have made use of realistic noise estimates provided by the PLATO project, which include estimated instrumental and other contributions beyond the expected shot noise levels. Moreover, that our proposed sample contains bright targets lends confidence that contamination from background sources will not be a serious issue for any analysis.

Finally, we note that an ensemble approach may be useful in other scenarios. Stellar clusters are an obvious example, targeting ranges of the cluster isochrone that lie beyond the range of detectability for individual stars. The cumulative gains given by combining data on different stars are much easier to predict in the cluster scenario, since all member stars lie at essentially the same distance, unlike in the field star case. The proposed ESA HAYDN mission \citep{Miglio2024} would provide an excellent opportunity to exploit the approach.

\begin{acknowledgements}

This work presents results from the European Space Agency (ESA) space mission PLATO. The PLATO payload, the PLATO Ground Segment and PLATO data processing are joint developments of ESA and the PLATO mission consortium (PMC). Funding for the PMC is provided at national levels, in particular by countries participating in the PLATO Multilateral Agreement (Austria, Belgium, Czech Republic, Denmark, France, Germany, Italy, Netherlands, Portugal, Spain, Sweden, Switzerland, Norway, and United Kingdom) and institutions from Brazil. Members of the PLATO Consortium can be found at \url{https://platomission.com/}. The ESA PLATO mission website is \url{https://www.cosmos.esa.int/plato}. We thank all teams working for PLATO. W.J.C., G.R.D, M.B.N. and A.S. acknowledge the support of the UK Space Agency. T.L.C. is supported by Funda\c c\~ao para a Ci\^encia e a Tecnologia (FCT) in the form of a work contract (\href{ https://doi.org/10.54499/2023.08117.CEECIND/CP2839/CT0004}{2023.08117.CEECIND/CP2839/CT0004}). M.N.L. acknowledges support from the ESA PRODEX programme (PEA 4000142995). This research has made use of NASA's Astrophysics Data System Bibliographic Services.

\end{acknowledgements}

\bibliographystyle{aa} 
\bibliography{aa59600-26.bib} 

\begin{appendix}
\onecolumn
\nolinenumbers

\section{Estimation of equivalent stellar number densities for PLATO sample}
\label{sec:num}

\FloatBarrier
\begin{figure*}
\centering
\centerline{\includegraphics[width=0.45\textwidth]{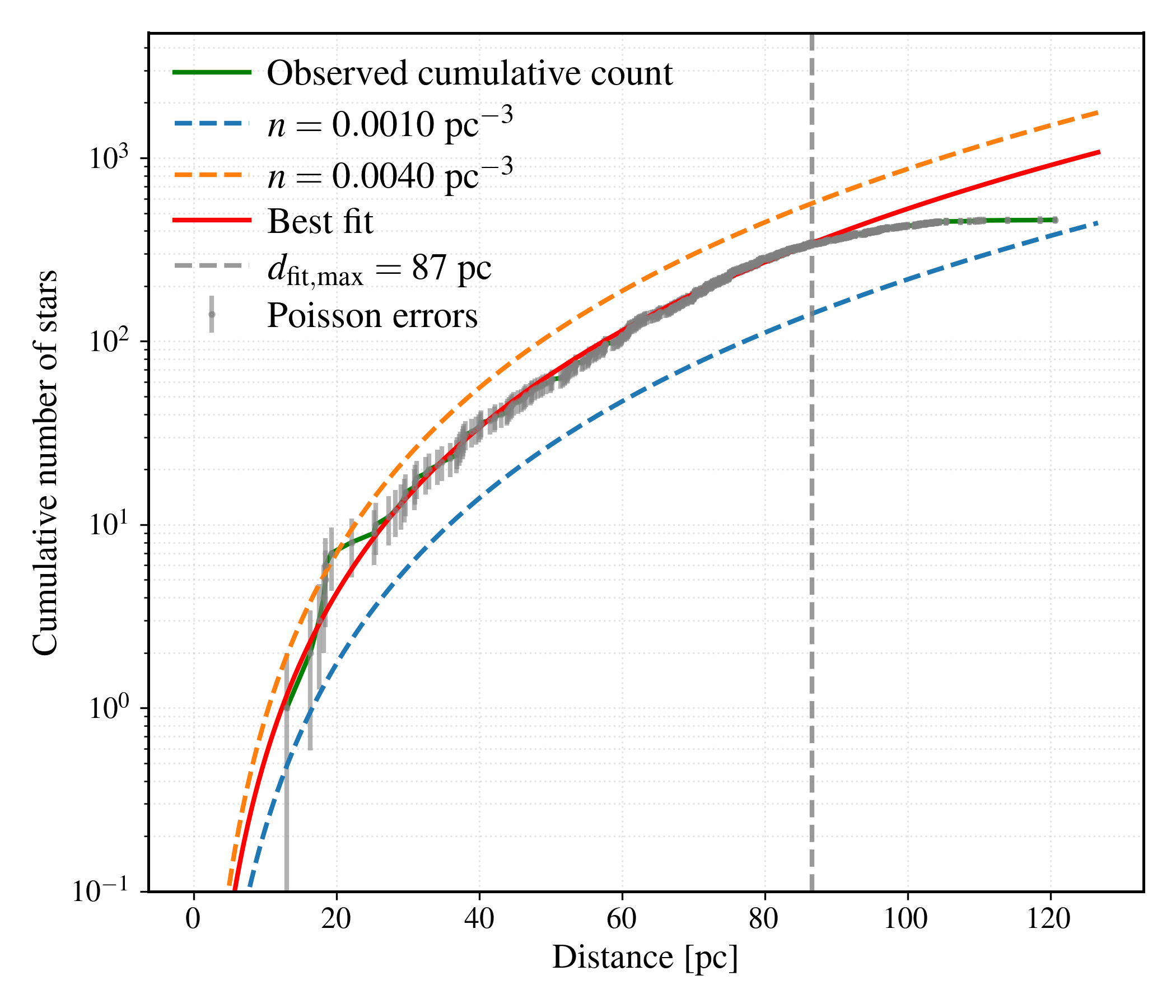}
            \includegraphics[width=0.45\textwidth]{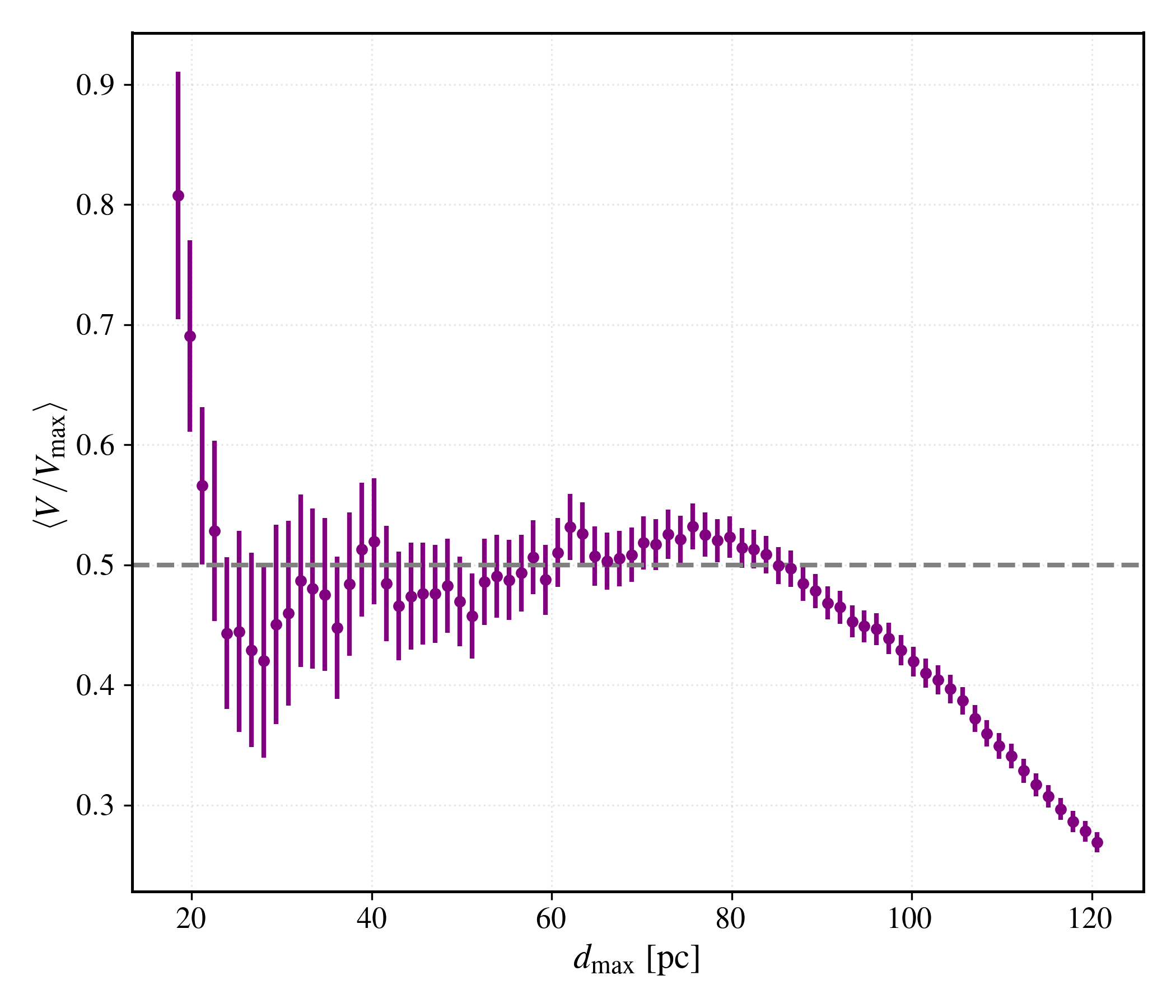}}
        \caption{Left: Cumulative counts versus distance for the sample in the temperature bin $4300 \leq T_{\rm eff} \leq 4700\,\rm K$, with Poisson error bars. The red line shows the best-fitting model, assuming a volume-complete sample out to $d_{\rm fit,max} = 87\,\rm pc$. That limit was fixed by the results shown in the right-hand panel, from performing a Schmidt uniformity test on the selected sample as a function of distance (see text for details).}
          \label{fig:num4347}
\end{figure*}

The left-hand panel of Fig.~\ref{fig:num4347} shows the cumulative count of selected stars that lie the range $4300 \leq T_{\rm eff} < 4700\,\rm K$, plotted as a function of distance and with Poisson error bars. The red line is a best-fit model of the number density $n$, assuming the sample is consistent with being uniformly distributed and volume complete out to some distance $d_{\rm fit,max}$. Before coming back to the fit, we first discuss how we determined $d_{\rm fit,max}$ by performing a uniformity test (\citealt{Schmidt1968}) on the sample (see also \citealt{Prisinzano2026}). Consider those stars that lie within a distance $d_{\rm max}$ that encloses the volume $V_{\rm max} = 4/3\pi d_{\rm max}^3$. Each of the $N_{\rm max}$ stars within that volume lie at some distance $r_i$, which encloses the volume $V_i = 4/3\pi r_i^3$. In a uniformly distributed sample, the sum
 \begin{equation}
 \left< V / V_{\rm max} \right> = N_{\rm max}^{-1} \sum_i V_i/V_{\rm max}
 \end{equation}
has an expectation value of 0.5, and a standard error of $(12N_{\rm max})^{-1/2}$. The right-hand panel of Fig.~\ref{fig:num4347} plots this statistic for a range of assumed $d_{\rm max}$. The sample is seen to deviate from 0.5 by more than one standard error at a distance of $d_{\rm fit,max} = 87\,\rm pc$. The red line in the left-hand panel then shows the result of fitting the cumulative star counts to a simple model having a constant number density $n$, i.e. $V(r) = 4/3 \alpha n \pi r^3$, out to the distance $d_{\rm fit,max}$. With a PLATO all-sky fraction of $\alpha = 0.052$, the fit returns a best-fitting stellar density of $n \simeq 2.4 \times 10^{-3}\,\rm pc^{-3}$. This value is not out of line with results from the RECONS survey\footnote{REsearch Consortium On Nearby Stars, www.recons.org.} (e.g. see \citealt{Henry1994,Henry2024}), and other studies using those data (e.g. see \citealt{Winters2019}).

\section{Construction of artificial PLATO power spectra}
\label{sec:artspec}

We used the individual data on each star in the PIC as input to our computation of the artificial PLATO power spectra. The required input data included the seismic parameters $\nu_{\rm max}$ (discussed in Sect.~\ref{sec:detect}),  $H_{\rm env}$, $\Delta\nu$, $A_{\rm max}$ and $\Gamma_{\rm env}$ (all discussed in Sect.~\ref{sec:oscpow}); and the background parameters $b_{\rm instr}$ and $b_{\rm gran}$ (discussed in Sect.~\ref{sec:backpow}). Below we describe how spectra were computed for individual stars. We then followed the procedures outlined in the main body of the paper to combine those individual spectra to give the composite, weighted spectra shown in Fig.~\ref{fig:artspec} for the $4300 \leq T_{\rm eff} < 4700\,\rm K$ temperature bin.

For a given star, we first laid down a set of mode frequencies using the asymptotic relation
 \begin{equation}
  \nu_{nl} = \Delta\nu \left( n + l/2 + \epsilon \right) - Dl(l+1).
\end{equation}
Here, $\nu_{nl}$ are the frequencies of modes of overtone number $n$ and angular degree $l$, and we include contributions for $5 \le n \le 30$ and $0 \le l \le 3$. The value of the phase term $\epsilon$ fixes the placement of the overtones in frequency. Since we are interested in detecting the weighted excess of power in the combined spectrum, the values chosen are not crucial. Nevertheless, guided by the results of \citet{White2011}---which suggest $\epsilon$ is primarily a function of $T_{\rm eff}$---we assume values for K dwarfs will be similar to those of red giants with detected oscillations that have similar $T_{\rm eff}$. We then drew $\epsilon$ for each star from a uniform random distribution, such that
\[
\epsilon \in U[0.7,1.3].
\]
The term $D$ fixes the small separations in frequency between adjacent modes of angular degree $l=0$ and 2, and $l=1$ and 3. Again, the exact values adopted here are not crucial. Guided by computed frequencies of stellar evolutionary models (e.g. see \citealt{White2011}), we drew $D$ for each star from a uniform random distribution, such that
\[
D \in U[1.3,2.5].
\]

We modelled the rotationally split, azimuthal components of each non-radial mode according to
\begin{equation}
  \nu_{nlm} = \nu_{nl} + m \delta\nu_{\rm rot},
\end{equation}
where the rotational splitting parameter $\delta\nu_{\rm rot}$ is assumed to be the same for all modes (e.g. see \citealt{Ball2018}). As per $\epsilon$ and $D$, the exact values adopted for the splitting are not critical. Here, we used the results of \citet{Santos2019} on measured surface rotation periods $P_{\rm rot}$ of K dwarfs observed by \emph{Kepler}, taking their median period of $P_{\rm rot} \simeq 25\,\rm d$ to fix the splitting for all stars at a value of $\delta\nu_{\rm rot} = 1/P_{\rm rot} = 0.46\,\rm \mu Hz$.

The height (maximum power spectral density) of each mode in the power spectrum is given by
 \begin{equation}
  H_{nlm} = \frac{2V_l \varepsilon_{lm}(i) A_{\rm max}^2}{\pi \Gamma_{nl}}  \exp \left[ -\,4\ln 2 \left( \frac{\nu_{nlm} - \nu_{\max}}{\Gamma_{\rm env}} \right)^2 \right].
 \label{eq:Hnlm}
 \end{equation}
The term $V_l$ fixes the relative mode visibilities for different angular degrees, where $\varsigma$ (see Sect.~\ref{sec:oscpow}) corresponds to the sum of these visibilities. Following the computations for PLATO in \citet{Lund2026}, we adopted values of $V_0 = 1.0$, $V_1=1.54$, $V_2=0.51$ and $V_3=0.10$. The term $\varepsilon_{lm}(i)$ fixes how the visibility of each mode component is affected by the stellar angle of inclination $i$ presented by the star, and is calculated from (\citealt{Gizon2003}) \begin{equation}
   \varepsilon_{lm}(i) = \frac{(l-|m|)!}{(l+|m|)!} [P_l^{|m|} (\cos i)]^2. 
 \end{equation}
Here, we assumed the orientations in space of the rotation axes of our sample of stars are random, meaning we could assign an angle $i$ for each star according to (e.g. see \citealt{Ball2018})
 \[
 i = \cos^{-1} (U[0,1]).
 \]
The variable $\Gamma_{nl}$ in Eq.~\ref{eq:Hnlm} fixes the widths of the mode peaks in the frequency domain. These widths are known to vary with frequency in any given solar-like oscillator (e.g. see results on \emph{Kepler} stars in \citealt{Lund2017}). Capturing that variation is not critical to the outcome of these simulations, as per the reasons we have given above for other parameters. For the sake of simplicity we therefore adopted a single width for all modes in all stars. We expect the widths for K dwarfs to be narrower than those shown by detected solar-like oscillations in G dwarfs. Here, we fixed $\Gamma_{nl} = 0.25\,\rm \mu Hz$. 

With all oscillation parameters computed, the total underlying noise-free (limit) spectrum due to the modes is given by
 \begin{equation}
 P_{\rm osc} = \sum_{nlm} H_{nlm} \left[ 1 + \left( \frac{\nu - \nu_{nlm}}{\Gamma_{nl}/2}\right)^2 \right]^{-1}.
 \end{equation}
The full limit spectrum then includes contributions from granulation and instrumental noise, and is given by
 \begin{equation}
 P(\nu) = \eta^2[P_{\rm osc}(\nu)+ P_{\rm gran}(\nu)] + b_{\rm instr},
 \end{equation}
where for completeness we include the factor $\eta = {\rm sinc}[\pi(\nu \Delta t)]$ that captures the signal attenuation due to the finite sampling $\Delta t$, though as noted in Sect.~\ref{sec:backpow} the cadence of 25\,s means this has a negligible impact in the range of frequencies of interest here. 

The power spectral density due to granulation is calculated from 
 \begin{equation}
 P_{\rm gran}(\nu)  = \frac{b_{\rm gran}[1 + (2\pi \nu_{\rm max} \tau_{\rm gran})^2]}{1+ (2\pi \nu \tau_{\rm gran})^2},
 \end{equation}
where we recall that $b_{\rm gran}$ corresponds to the granulation power spectral density at $\nu_{\rm max}$ (see Sect.~\ref{sec:backpow}), and the granulation timescale is given by
 \begin{equation}
 \tau_{\rm gran} = \tau_{\rm gran,\odot} \, (\nu_{\rm max,\odot} / \nu_{\rm max}),
 \end{equation}
and we adopt $\tau_{\rm gran,\odot} = 200\,\rm sec$.

The grid of frequencies $\nu$ on which the spectrum is calculated is given by $\nu = j/T$, where the dummy integer $j$ runs from $0 \le j \le \nu_{\rm Nyq}/\Delta_T$. Here $\nu_{\rm Nyq} \equiv 1/(2\Delta t) = 20,000\,\rm \mu Hz$ is the Nyquist frequency for the PLATO cadence of $\Delta t = 25\,\rm sec$ (see also Sect.~\ref{sec:backpow}); and $\Delta_T = 1/T$ (see Sect.~\ref{sec:comb1}) corresponds to the frequency resolution at critical sampling. With the underlying noise-free power spectrum computed, the final step added stochastic noise---commensurate with the required $\chi^2$ 2-d.o.f. statistics---to give a proxy of the observed spectrum $P^{\rm obs}(\nu)$ of the star. This was achieved by the following computation (\citealt{Duvall1990}):
\begin{equation}
P^{\rm obs}(\nu) = - P(\nu) \ln ( U[0,1]).
\end{equation}

\end{appendix}
\end{document}